\documentclass{article}
\usepackage[preprint]{neurips_2026}

\usepackage[utf8]{inputenc}
\usepackage[T1]{fontenc}
\usepackage{hyperref}
\usepackage{url}
\usepackage{xurl}
\usepackage{booktabs}
\usepackage{graphicx}
\usepackage{wrapfig}
\usepackage{amsmath}
\usepackage{float}
\usepackage{amsfonts}
\usepackage{nicefrac}
\usepackage{microtype}
\usepackage{xcolor}
\usepackage[most]{tcolorbox}
\usepackage{array}
\usepackage{tabularx}
\usepackage{multirow}
\usepackage{titletoc}
\usepackage{makecell}
\usepackage{tikz}
\usetikzlibrary{positioning,shapes.geometric,arrows.meta,fit,calc,backgrounds,decorations.pathreplacing}

\IfFileExists{fontawesome5.sty}{%
  \usepackage{fontawesome5}
  \newcommand{\githubicon}{\faGithub}
}{%
  \newcommand{\githubicon}{%
    \tikz[baseline=-0.55ex,scale=0.12]{%
      \fill[black] (0,0) circle (1);
      \fill[white] (-0.32,0.18) circle (0.16);
      \fill[white] (0.32,0.18) circle (0.16);
      \fill[white] (-0.45,0.52) -- (-0.62,0.95) -- (-0.12,0.70) -- cycle;
      \fill[white] (0.45,0.52) -- (0.62,0.95) -- (0.12,0.70) -- cycle;
      \fill[white] (-0.50,-0.18) .. controls (-0.42,-0.62) and (0.42,-0.62) .. (0.50,-0.18)
        .. controls (0.42,-0.42) and (-0.42,-0.42) .. (-0.50,-0.18);
    }%
  }
}

\newcolumntype{P}[1]{>{\raggedright\arraybackslash}p{#1}}
\newcolumntype{C}[1]{>{\centering\arraybackslash}m{#1}}
\newcolumntype{Y}{>{\raggedright\arraybackslash}X}

\titlecontents{section}[1.5em]{}{\contentslabel{1.5em}}{\hspace*{-1.5em}}{\titlerule*[0.5pc]{.}\contentspage}
\titlecontents{subsection}[3em]{}{\contentslabel{2.3em}}{\hspace*{-2.3em}}{\titlerule*[0.5pc]{.}\contentspage}

\newtcolorbox{NewBoxFloat}[2]{%
  enhanced,
  colback=white,
  colframe=black!65,
  colbacktitle=gray!12,
  coltitle=black,
  title={#1},
  fonttitle=\bfseries\small,
  boxrule=0.5pt,
  arc=1.2mm,
  left=2.5mm, right=2.5mm,
  top=1.4mm, bottom=1.4mm,
  titlerule=0pt,
  label={#2},
}

\definecolor{midnightgreen}{rgb}{0.0, 0.29, 0.33}

\AtBeginEnvironment{thebibliography}{\sloppy}
\newcommand{\repoURL}{https://github.com/cxcscmu/Auto-Research-Recipes}
\newcommand{\paperrepository}{%
  \begin{center}
    \small \href{\repoURL}{\githubicon\ GitHub Repository}
  \end{center}
}

\title{Auto Research with Specialist Agents Develops Effective and Non-Trivial Training Recipes}

\author{%
  Jingjie Ning \quad Xiaochuan Li \quad Ji Zeng \quad Hao Kang \quad Chenyan Xiong\\
  School of Computer Science, Carnegie Mellon University\\
  \texttt{\{jening, xiaochu4, jizeng, haok, cx\}@cs.cmu.edu}
}

\begin{document}
\maketitle
\paperrepository

\begin{abstract}
We study auto research as a closed empirical loop driven by external measurement. Each submitted trial carries a hypothesis, an executable code edit, an evaluator-owned outcome, and feedback that shapes the next proposal. The output is not a generated paper or a single model checkpoint, but an auditable trajectory of proposals, code diffs, experiments, scores, and failure labels. We instantiate this loop with specialist agents that partition recipe surfaces and share measured lineage across trials. The central empirical finding is that lineage feedback lets agents turn evaluator outcomes, including crashes, budget overruns, size failures, and accuracy-gate misses, into later program-level recipe edits rather than one-shot suggestions. Across 1,197 headline-run trials plus 600 Parameter Golf control trials after one-time setup and launch, humans did not choose proposals, edit recipes, override scores, or repair failed trials during the search. In the three headline runs, the same submitted-trial loop reduces Parameter Golf validation bpb by $0.81\%$, raises NanoChat-D12 CORE by $38.7\%$, and reduces CIFAR-10 Airbench96 wallclock by $4.59\%$, with each task measured by its own external evaluator and legality checks. The trace includes a strict architecture-domain audit of 157 headline-run submissions and program rewrites such as a NanoChat attention-kernel path change. Within this scope the loop autonomously writes code, submits experiments, absorbs feedback, applies and combines known techniques inside each environment, and improves public starting recipes.
\end{abstract}

\section{Introduction}
\label{sec:intro}

Machine learning research advances by measured iteration: change code, launch experiments, read results, and choose the next move. This paper hands that propose-measure-revise loop to language agents under the same measurement environment a human researcher would use. Here, auto research means agents propose hypotheses, edit code, submit experiments, read evaluator-owned outcomes, and use them to revise later proposals. After one-time setup and launch, humans do not choose trials during search. Its unit is a submitted trial rather than a generated narrative: a hypothesis, executable code edit, evaluator-owned outcome, and feedback signal. The channel records successes and failures as measured evidence rather than polished summaries.

Training recipes are a natural testbed because they expose architecture, data, optimization, schedules, losses, compression, and systems under constraints. An edit can improve quality but exceed a size cap, save time but miss an accuracy gate, or expose a bottleneck convertible into training tokens. These feedback shapes make the loop follow measured evidence rather than a fixed grid. Lineage is the cross-trial record of hypotheses, diffs, scores, runtimes, statuses, and crash summaries read before the next proposal. Specialist roles partition the recipe surface, while shared lineage carries measured evidence across roles so neighboring surfaces can build on it.

Prior work establishes pieces of this picture across repository-editing agents, machine-learning experiment agents, and evaluator-driven discovery systems. We study the empirical regime where these pieces form a sustained feedback loop over real training recipes, with executable edits, external measurements, failures, and follow-up proposals analyzed as one measured artifact.

We study three environments with complementary feedback. Parameter Golf exposes size and budget pressure under a fixed FineWeb loss task~\citep{openai2025parametergolf,penedo2024fineweb}; NanoChat-D12 exposes wallclock headroom and runtime bottlenecks in fixed-budget pretraining~\citep{karpathy2025nanochat,li2024datacomplm}; and CIFAR-10 Airbench96 exposes an accuracy gate around speed improvements~\citep{jordan2024airbench,krizhevsky2009cifar}. Across the headline runs, the same loop improves all three starting recipes, and the traces expose auto research through code edits, launched runs, measurements, crashes, and follow-up proposals. The empirical object is the trajectory after successful and failed trials. The loop writes code, launches experiments, reads evaluator-owned outcomes, and uses feedback to revise later proposals. It improves public starting recipes, applies known techniques, and runs without human intervention during search. In these trials, agents combine and transfer known techniques rather than propose anything as structurally novel as the original Transformer.

\begin{figure}[!t]
    \centering
    \includegraphics[width=\linewidth]{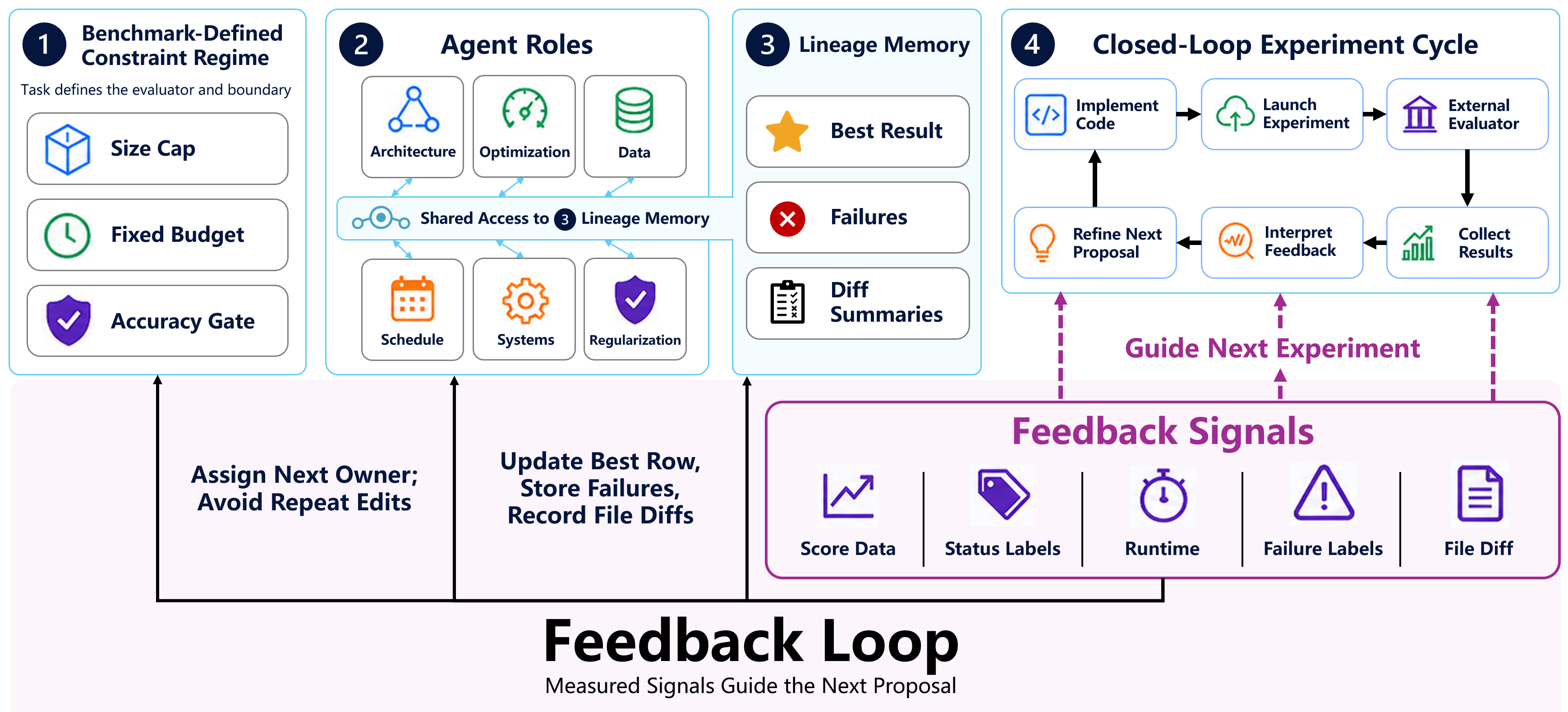}
    \vspace{-0.8\baselineskip}
    \caption{Closed-loop auto research trajectory. Submitted trials connect proposals, executable edits, external measurements, feedback, and the next research move.}
    \label{fig:overview}
    \vspace{-0.8\baselineskip}
\end{figure}

The contributions are to formulate auto research as an auditable closed-loop trajectory rather than a single generated output, instantiate it in compute-budgeted training-recipe development, demonstrate autonomous externally measured research without human intervention inside the search loop, and analyze measured lineage, program-level edits, failure feedback, evaluator-owned measurement, and role-partitioned recipe search. In a representative NanoChat-D12 trace, a systems agent diagnosed an attention-backend bottleneck. Recovered wallclock returned through lineage as budget headroom, later proposals spent it on more tokens, and the improved CORE score became the next current best. Figure~\ref{fig:overview} summarizes how proposals, code edits, external measurements, and lineage feedback become the next research move.

\section{Related Work}
\label{sec:related}

\paragraph{Evaluator-driven program search and parameter optimization.}
AlphaDev, FunSearch, and AutoML-Zero propose programs and let an evaluator decide validity~\citep{mankowitz2023alphadev,romera2024funsearch,real2020automlzero}. AlphaEvolve extends this to an evolutionary coding agent under automated evaluator feedback, but still targets algorithms and infrastructure rather than full training recipes~\citep{novikov2025alphaevolve}. Hyperparameter optimization, population based training, and neural architecture search also use measured selection, usually over fixed parameter or architecture spaces~\citep{bergstra2012random,snoek2012bayesopt,li2018hyperband,jaderberg2017pbt,zoph2017nas,real2019regularized,liu2019darts}. We keep the evaluator-driven pattern and move it to full Python training pipelines with data loading, optimizer state, schedules, kernels, evaluation, and legality checks, where crashes, artifact caps, and runtime bottlenecks become feedback and the measured trajectory is analyzed, not only the final score.

\paragraph{Language agents for code, machine learning, and long-running tasks.}
SWE-bench and SWE-agent test repository editing and agent-computer interfaces~\citep{jimenez2024swebench,yang2024sweagent}, while MLAgentBench and MLE-bench move agents into repeated ML experiments~\citep{huang2023mlagentbench,chan2024mlebench}. RE-Bench evaluates open-ended ML research engineering against human experts~\citep{wijk2025rebench}; MLGym-Bench frames open-ended AI research as agent environments and finds gains often come from hyperparameters rather than new hypotheses, algorithms, or architectures~\citep{nathani2025mlgym}; AIBuildAI studies hierarchical model-building agents on MLE-Bench~\citep{zhang2026aibuildai}; and PostTrainBench asks frontier agents to improve LLM post-training under bounded compute while exposing reward-hacking failures~\citep{rank2026posttrainbench}. The AI Scientist adds idea generation and paper writing~\citep{lu2024aiscientist}, and Anthropic reports on effective, multi-agent, and long-running coding agents provide practical context~\citep{anthropic2024buildingagents,anthropic2025multiagentresearch,anthropic2025longrunningagents}. Our bounded setting instead makes the output a measured trajectory of code edits on fixed training tasks, so the closed empirical loop itself is the object of study.

\paragraph{Compute-budgeted training and efficient training tools.}
Compute-optimal training studies how model size, data, and compute scale~\citep{hoffmann2022chinchilla}; nanoGPT, nanochat, Parameter Golf, and CIFAR-10 Airbench make related tradeoffs runnable at smaller scale~\citep{karpathy2023nanogpt,karpathy2025nanochat,openai2025parametergolf,jordan2024airbench}. Parameter Golf uses a FineWeb-derived slice with artifact and wallclock limits~\citep{penedo2024fineweb}, nanochat provides an end-to-end language-model pipeline with CORE-style evaluation from DataComp-LM~\citep{li2024datacomplm}, and Airbench provides fast CIFAR-10 recipes with explicit accuracy and time targets~\citep{krizhevsky2009cifar}. Final recipes often reuse tools such as FlashAttention and GPTQ~\citep{dao2022flashattention,frantar2023gptq}. These tasks are cheap enough for repeated calls but strict enough to reject shortcuts, testing whether agents can choose and combine known tools under budgets without humans selecting the next trial.

\section{Closed-Loop Auto Research Methodology}
\label{sec:methodology}
\label{sec:benchmarks}
\label{sec:agent}

The method pairs externally measured training-recipe environments with a submitted-trial feedback loop. The environment fixes editable files, the scored metric, legal failures, and evaluator feedback. The loop turns that feedback into later hypotheses and code edits. The four levels are task feedback, submitted trials, shared lineage, and parallel iteration.

\subsection{Task environments and feedback signals}

We use three environments because they expose different feedback through the same submitted-trial loop. Parameter Golf rewards lower validation bits per byte on a fixed FineWeb-derived task with a 16 MB artifact cap and a 10 minute budget on eight H100 GPUs~\citep{openai2025parametergolf,penedo2024fineweb}. We use the public 1.0810 leaderboard score as the denominator, keeping the delta tied to the public target record. Each trial returns score, status, exact byte counts, and per-phase timing, so the dominant feedback is size and budget pressure around the current bpb frontier.

NanoChat-D12 rewards higher CORE from a fixed d12 nanochat pretraining run~\citep{karpathy2025nanochat,li2024datacomplm}. The starting point is one calibrated run of the unmodified upstream recipe at the pinned commit, reaching 0.1618 CORE in our GPU environment. Agents can edit the coordinator script and vendored nanochat Python tree, but trials cannot download during execution. Tokenizer files, pretraining shards, and the evaluation bundle are prepared before launch. The protected parser extracts CORE from the log, and the main feedback is wallclock headroom under the fixed budget, because faster code can spend recovered time on more tokens.

CIFAR-10 Airbench96 rewards lower shell-measured wallclock time, but only when mean CIFAR-10 accuracy reaches at least 0.96~\citep{jordan2024airbench,krizhevsky2009cifar}. The starting point is the unmodified Airbench96 recipe calibrated to 26.356~s under our ten-seed cold-process protocol. The recipe cannot report its own time: the run script writes timing sidecars, and the classifier reads them. The main feedback comes from the accuracy gate, where fast near-misses return timing plus accuracy rather than a generic crash, making the miss usable for the next proposal. In all three environments, the starting recipe is fixed before search and the editable recipe does not own the evaluator. For each frozen run, the harness, prompt templates, static knowledge files, and specialist taxonomy are fixed before launch; no human intervention occurs during that reported trajectory.

\subsection{Submitted-trial loop}

A trial is the unit of the empirical loop. The task fixes editable files, score field, legality checks, and submission path. An agent reads current lineage, proposes a hypothesis, implements it as executable code, and submits a trial. An external evaluator measures the run, assigns status, and appends score, timing, and failure information. The next agent receives this feedback and refines the next proposal.

Each agent session is a bounded LLM-agent SDK call, not an always-running process. It receives a fresh lineage view at session start, may submit multiple trials when a result exposes a concrete follow-up edit, and terminates under a tool-turn cap. All scores are measured outside the editable recipe. Parameter Golf uses the official evaluation path. NanoChat-D12 uses a protected parser and evaluator-side classifier path, with edits audited for parser or evaluator touches. CIFAR uses shell-side timing and rejects trials that miss the accuracy gate. This prevents reward hacking such as printing a better score or reporting fake runtime.

\subsection{Specialist roles and shared lineage}

Specialist roles partition the editable recipe surface by environment constraints. The taxonomy is chosen before each run and fixed during search. Section~\ref{sec:feedback-controls} compares this role decomposition against generic multi-agent and single-agent controls. Parameter Golf has a broad recipe surface under a hard artifact cap, so its ten specialists cover architecture, optimization, quantization, regularization, loss, evaluation, curriculum, tokenizer, test-time training, and meta search. NanoChat-D12 is fixed-budget pretraining, so its five specialists cover architecture, optimization, data, schedule, and systems. CIFAR-10 Airbench96 is an accuracy-gated speed task, so its five specialists cover architecture, optimization, augmentation, loss, and regularization.

Each specialist sees the same metric but receives a different recipe-surface prompt. This role conditioning makes sessions attend to different surfaces rather than repeatedly editing the most salient knob. The run log stores hypothesis text, diff summary, score, status, timing, and crash reason. The prompt renderer selects a compact lineage slice for the next trial, including the current best row, specialist recent rows, and adjacent-specialist rows. This preserves the frontier and keeps failed directions visible without replaying the full transcript. This setup also makes the research process a releasable artifact. Each trial has a proposal summary, code-diff summary, measured score, status label, timing record, and failure summary when applicable. These traces do not rely on private model internals, so they can be released with the harness and final recipes for audit, reproduction, and follow-up analysis. The public code and artifact archive is available at \url{https://github.com/cxcscmu/Auto-Research-Recipes}.

\subsection{Measurement, calibration, and affordable iteration}

When hardware or run protocol differs from a public number, calibration runs before search and is append-only. This preserves logs and avoids stale denominators for NanoChat-D12 and CIFAR-10 Airbench96. Affordable iteration is a condition for closed-loop research because outcomes must return quickly enough to shape later proposals within the same search horizon. Our environments meet this condition because expensive phases are capped or short, while parallel submissions, score parsing, status classification, and legality checks run outside the editable recipe.

For environment $e$, write the continuous wallclock for one submitted trial as a run, evaluation, queue, and logging decomposition. With $N$ independent submitters using one shared blackboard, the measured throughput is
\begin{equation}
\tau_e=\tau^{\mathrm{run}}_e+\tau^{\mathrm{eval}}_e+\tau^{\mathrm{queue}}_e+\tau^{\mathrm{log}}_e;\quad
T_e(N)=\frac{N\,\eta_{\parallel,e}}{\tau_e};\quad
\eta_{\parallel,e}=\frac{T_e(N)}{N\,T_e(1)}\in(0,1].
\label{eq:parallel-efficiency}
\end{equation}
We estimate this on Parameter Golf with the same starting recipe, 600 second budgets, and continuous wallclock only, excluding human pauses. Over the matched first-200-trial window, the single-generalist variant clears 2.26 trials per hour. The ten-specialist role swarm clears 18.15 trials per hour, giving $\eta_{\parallel,\mathrm{PG}}\approx0.80$ against the ideal $10\times$ speedup. The ten generic agents clear 16.79 trials per hour with $\eta_{\parallel,\mathrm{PG}}\approx0.74$. Thus the role-versus-generic difference in Section~\ref{sec:experiments} is mainly proposal diversity and boundary discipline, not raw throughput. Efficiency is below one because submitters share the GPU pool, cluster queue, and blackboard filelock. This throughput matters because feedback helps the next proposal only when enough outcomes arrive within the same search horizon.

\section{Experiments}
\label{sec:experiments}

The experiments treat end-of-run score as a prerequisite, not the only object. We test whether the loop runs autonomously while writing code, submitting experiments, and collecting feedback; whether it improves each environment; whether submitted proposals include program-level changes rather than only numeric knobs; how outcomes distribute across roles; and how Parameter Golf controls isolate organization and feedback memory. The three headline runs contain 1,197 submitted trials: 900 in Parameter Golf, 200 in NanoChat-D12, and 97 in CIFAR-10 Airbench96. The three additional Parameter Golf control runs in Section~\ref{sec:feedback-controls} add 600 independent trials from the same starting recipe; the role-swarm control row in Table~\ref{tab:main} is the first 200-trial window of the 900-trial headline run and is not counted again. Two historical 91-trial traces are retained only for proposal-diversity audit and excluded from these totals.

We operationalize the loop through the trial log. Each trial records the proposing role, edit domain, proposal and diff summaries, status, score delta when valid, failure type when invalid, and timing or crash metadata. This is the observed proposal surface, not the latent distribution over every considered idea. We analyze which code edits reached the evaluator and how feedback shaped the trajectory.

\subsection{Main trajectories}

\begin{figure}[t]
\centering
\includegraphics[width=1.0\linewidth]{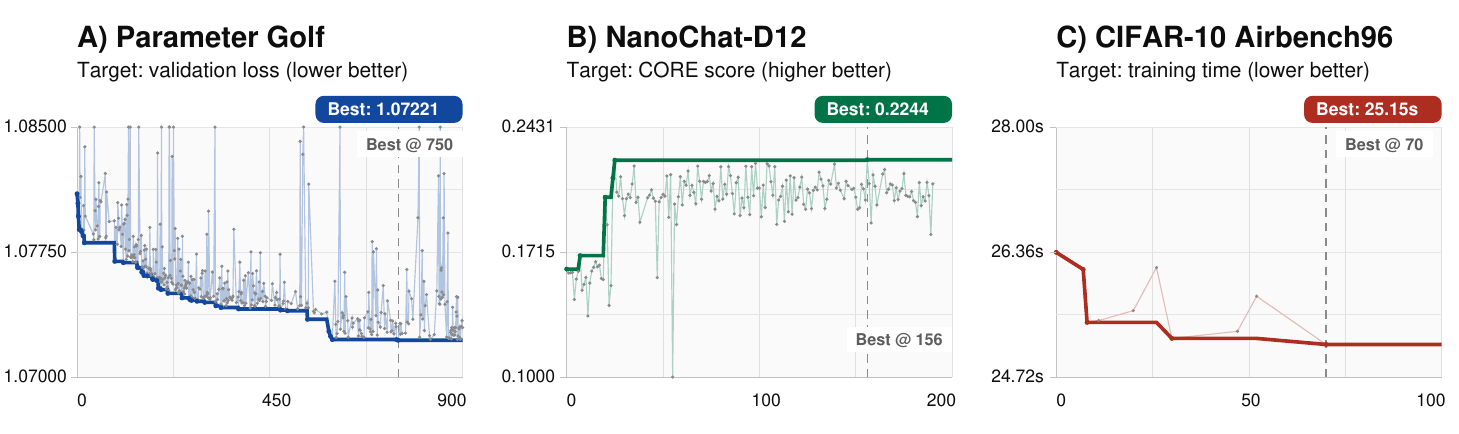}
\vspace{-1.0\baselineskip}
\caption{Best-so-far score over submitted trial index. Points are valid measured trials only; ineligible trials are omitted. The bold line shows best-so-far.}
\label{fig:trajectories}
\vspace{-0.8\baselineskip}
\end{figure}

\begin{table}[!t]
\centering
\footnotesize
\caption{Main experimental summary. Reference rows give external numbers, starting-point rows give the fixed search recipe and denominator for relative change, and run rows report change against that start. For Parameter Golf, 1.2244 is OpenAI's official naive task reference, while 1.0810 is the public starting recipe. Dashes mark rows without submitted trials. Scores are rounded to four decimals here; exact trace values appear where individual controls are discussed.}

\label{tab:main}
\setlength{\tabcolsep}{3pt}
\renewcommand{\arraystretch}{1.0}
\begin{tabularx}{\linewidth}{@{}
P{0.18\linewidth}
X
C{0.13\linewidth}
C{0.11\linewidth}
C{0.07\linewidth}
C{0.10\linewidth}
@{}}
\toprule
Environment & Row & Score & \makecell{Rel. vs\\start} & Trials & \makecell{Valid\\impr.} \\
\midrule

{} &
Naive reference baseline &
1.2244 &
-- &
-- &
-- \\
\cmidrule(l){2-6}
& Public SOTA starting point &
1.0810 &
-- &
-- &
-- \\
\cmidrule(l){2-6}
& Role swarm, full run &
\textbf{1.0722} &
\textbf{-0.81\%} &
\textbf{900} &
\textbf{36} \\
\cmidrule(l){2-6}
\raisebox{0pt}[0pt][0pt]{\makecell[l]{Parameter Golf\\{\scriptsize \texttt{val\_bpb} (lower better)}}} &
Role swarm control &
1.0731 &
-0.73\% &
200 &
16 \\
\cmidrule(l){2-6}
& Single generalist &
1.0754 &
-0.52\% &
200 &
14 \\
\cmidrule(l){2-6}
& Generic-10 &
1.0745 &
-0.60\% &
200 &
10 \\
\cmidrule(l){2-6}
& No lineage &
1.0774 &
-0.33\% &
200 &
3 \\

\midrule
\raisebox{-0.45\baselineskip}[0pt][0pt]{\makecell[l]{NanoChat-D12\\{\scriptsize CORE (higher better)}}} &
Calibrated upstream start &
0.1618 &
-- &
-- &
-- \\
\cmidrule(l){2-6}
& Role swarm &
\textbf{0.2244} &
\textbf{+38.7\%} &
\textbf{200} &
\textbf{5} \\

\midrule
{} &
\makecell[l]{Upstream reported reference{\scriptsize (same code)}} &
27.3000~s &
-- &
-- &
-- \\
\cmidrule(l){2-6}
\raisebox{0.35\baselineskip}[0pt][0pt]{\makecell[l]{CIFAR-10\\Airbench96\\{\scriptsize \texttt{train\_s} (lower better)}}} &
Calibrated upstream start &
26.3560~s &
-- &
-- &
-- \\
\cmidrule(l){2-6}
& Role swarm &
\textbf{25.1464~s} &
\textbf{-4.59\%} &
\textbf{97} &
\textbf{4} \\

\bottomrule
\end{tabularx}
\end{table}

All relative changes in Table~\ref{tab:main} use the search starting point, not every external reference. For NanoChat-D12, the calibrated upstream d12 recipe at 0.1618 CORE is both baseline and fixed search start, so the final 0.2244 gain uses only that denominator. For CIFAR-10 Airbench96, the 27.3000\,s reference and 26.3560\,s start are the same upstream recipe under different protocols, so agent improvement is computed from the calibrated start.

Table~\ref{tab:main} summarizes external references, starts, headline runs, and Parameter Golf controls. Table~\ref{tab:program-transforms} gives one compact representative per environment showing the loop is not only scalar recipe tuning, with the fuller list in Table~\ref{app:tab:program-transforms-full}. We audit submitted trials whose specialist or domain is architecture. This conservative, reproducible rule gives 95 of 900 Parameter Golf trials, 42 of 200 NanoChat-D12 trials, and 20 of 97 CIFAR trials, or 157 of 1,197 headline-run trials ($13.1\%$). The count includes crashes, discards, disqualifications, and valid improvements because it measures submitted ideas, not only final-best contributors. We use this $13.1\%$ as a strict lower-bound sanity check, not an estimate of the full non-scalar edit fraction, because systems, optimizer, and loss specialists sometimes rewrite executable structure, such as the NanoChat attention-kernel path. The rows give representative submitted transformations outside a fixed HPO space.

\begin{table}[t]
\centering
\footnotesize
\caption{Compact representative submitted program transformations. Rows include valid and failed trials because they summarize generated research ideas, not only final-best contributors.}
\label{tab:program-transforms}
\setlength{\tabcolsep}{4pt}
\renewcommand{\arraystretch}{1.0}

\begin{tabularx}{\textwidth}{@{}
    >{\raggedright\arraybackslash}p{0.15\textwidth}
    >{\raggedright\arraybackslash}p{0.16\textwidth}
    >{\raggedright\arraybackslash}X
@{}}
\toprule
Environment & Trial id(s) & Concrete architecture or program change \\
\midrule

Parameter Golf
& \texttt{245, 475, 538}
& Recurrent residual scaling; separate RoPE/NoPE query gains; per-head data-dependent attention-output gate. \\

\midrule
NanoChat-D12
& \texttt{007}
& \texttt{SSSL} to \texttt{L} attention path; masked SDPA math layers moved to Flash SDPA. \\

\midrule
CIFAR-10 Airbench96
& \texttt{040/044/053, 059/062}
& Residual-preserved ConvGroup depth reductions; wider-shallower blocks under the accuracy gate. \\

\bottomrule
\end{tabularx}
\vspace{-1.0\baselineskip}
\end{table}

Each submitted trial records a proposal, code edit, evaluator status, and feedback for later proposals. Across headline-run trials, the logs contain 45 keeps and 592 valid non-improvements, plus boundary feedback such as size blocks, budget overruns, crashes, and accuracy-gate disqualifications. These rows are not discarded attempts: the case studies show how size, runtime, and accuracy-gate feedback return as follow-up edits.

\begin{wrapfigure}[15]{r}{0.65\linewidth}
    \vspace{-0.8\baselineskip}
    \centering
    \includegraphics[width=\linewidth]{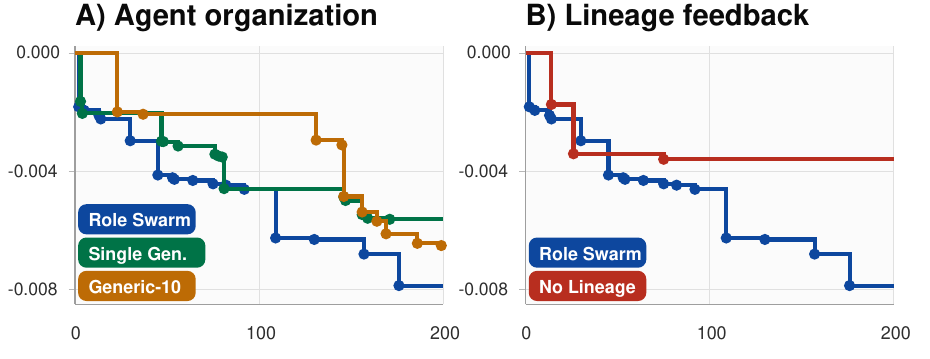}
    \vspace{-0.8\baselineskip}
    \caption{Parameter Golf controls over the first 200 trials. The y-axis is delta validation bpb, lower is better. Panel A compares agent organizations, and Panel B removes shared lineage.}
    \label{fig:pg-ablation}
    \vspace{-1.2\baselineskip}
\end{wrapfigure}

Figure~\ref{fig:trajectories} shows best-so-far score over submitted trial index using only valid measured points, including valid improvements and non-improvements. Ineligible trials are excluded. Earlier harness-vintage 91-trial traces, not prefixes of the 900-trial headline run, are retained in Table~\ref{tab:proposal-entropy} as historical proposal-diversity audits. Figure~\ref{fig:pg-ablation} is the primary Parameter Golf control because it shares the modern harness vintage and adds generic multi-agent and no-lineage controls.

\subsection{Loop behavior across roles}

Role-level outcomes provide trace context, but the submitted idea stream is primary. Appendix~\ref{app:trace-stats} reports role profiles, allocation balance, and tool-use summaries. The main text focuses on whether role-partitioned search changes the proposal surface and whether shared lineage carries ideas across role boundaries.

\paragraph{Proposal entropy and idea sharing.}

We audit the submitted idea stream directly. For each trial, we embed only recorded hypothesis text with TF-IDF, excluding role names, domains, scores, statuses, and implementation notes. We cluster proposals online: a proposal joins the nearest centroid if cosine similarity is at least 0.30, otherwise it starts a new cluster. The effective proposal count is $\exp(H)$, where $H$ is Shannon entropy over cluster sizes. This does not recover unsubmitted latent ideas, but measures how diverse evaluator-facing ideas were.

Table~\ref{tab:proposal-entropy} reports matched Parameter Golf controls and keeps the historical first-91 rows as a compact proposal-diversity audit. In the historical harness-vintage traces, the single generalist has 39.3 effective clusters and a 10.0\% near-duplicate rate, while the specialist swarm has 74.1 effective clusters and 0.0\% near duplicates under the same TF-IDF vocabulary. The contexts column records proposal partitions across role or agent contexts, with maximum rows per context in parentheses.

The same audit exposes idea sharing through lineage. In the historical 91-trial Parameter Golf swarm trace, 76 of 86 within-window parent edges cross role boundaries. Of the 7 keeps in that window, the 4 with within-window parents all build on another role's row. In the matched 200-trial controls, the role-decomposed lineage swarm has 10 of 12 successful keep parent edges crossing contexts, while no-lineage collapses to 0 of 1. The generic 10-agent control shows parallel contexts alone are not enough: it has 10 contexts and many cross-agent parent edges, but only 41.1 effective clusters because identical prompts concentrate the stream.

\begin{table}[t]
\centering
\scriptsize
\caption{Submitted-proposal entropy and information sharing in Parameter Golf controls.}
\label{tab:proposal-entropy}
\setlength{\tabcolsep}{2.4pt}
\renewcommand{\arraystretch}{0.96}

{
\renewcommand{\tabularxcolumn}[1]{>{\raggedright\arraybackslash}m{#1}}
\begin{tabularx}{\textwidth}{@{}
    >{\raggedright\arraybackslash}m{0.165\textwidth}
    >{\centering\arraybackslash}m{0.038\textwidth}
    >{\centering\arraybackslash}m{0.072\textwidth}
    >{\centering\arraybackslash}m{0.072\textwidth}
    >{\centering\arraybackslash}m{0.070\textwidth}
    >{\centering\arraybackslash}m{0.075\textwidth}
    >{\centering\arraybackslash}m{0.085\textwidth}
    >{\centering\arraybackslash}m{0.085\textwidth}
    X
@{}}
\toprule
Trace
& Rows
& Ctx.
& \makecell[c]{Eff.\\clusters}
& \makecell[c]{Top\\cluster}
& \makecell[c]{Near\\dup.}
& \makecell[c]{Cross-ctx.\\parent}
& \makecell[c]{Cross-ctx.\\keep}
& Shared idea clusters and limits \\
\midrule

Role swarm + lineage
& 200
& 10 (22)
& 134.8
& 3.5\%
& 2.0\%
& 154/184 (83.7\%)
& 10/12 (83.3\%)
& 5 of 29 clusters, 19 rows \\

Role swarm, no lineage
& 200
& 10 (27)
& 121.7
& 2.5\%
& 2.0\%
& 155/174 (89.1\%)
& 0/1 (0.0\%)
& 5 of 28 clusters, 19 rows; parent IDs lack rich lineage feedback \\

Generic 10-agent
& 200
& 10 (22)
& 41.1
& 12.0\%
& 1.5\%
& 125/158 (79.1\%)
& 7/9 (77.8\%)
& 22 of 30 clusters, 135 rows \\

\cmidrule(lr){1-9}

Single generalist
& 200
& 1 (200)
& 61.9
& 17.5\%
& 10.1\%
& \textit{n/a}
& \textit{n/a}
& \textit{n/a} \\

Historical swarm
& 91
& 10 (11)
& 74.1
& 3.3\%
& 0.0\%
& 76/86 (88.4\%)
& 4/4 (100.0\%)
& 5 of 8 clusters, 13 rows \\

Historical single
& 91
& 1 (91)
& 39.3
& 7.7\%
& 10.0\%
& \textit{n/a}
& \textit{n/a}
& \textit{n/a} \\

\cmidrule(lr){1-9}

Role swarm (full run)
& 900
& 10 (96)
& 439.6
& 2.2\%
& 1.1\%
& 781/895 (87.3\%)
& 32/33 (97.0\%)
& 31 of 134 clusters, 168 rows \\

\bottomrule
\end{tabularx}

\vspace{0.5pt}
\begin{minipage}{0.98\textwidth}
\footnotesize
\emph{Notes.} Clusters use hypothesis text only; effective clusters are $\exp(H)$ at cosine threshold 0.30. Top cluster is the largest row share. Contexts are role or agent partitions, with max rows in parentheses. Parent and keep fractions report cross-context edges. Keep denominators include successful improvements whose declared parent falls inside the audited window. In the no-lineage run, declared parent IDs remain supervisor bookkeeping and rebasing anchors; because agents receive no within-run prior-trial content beyond the current-best score line, these edges are ancestry rather than information transfer.
\end{minipage}
}
\vspace{-1.0\baselineskip}
\end{table}

\subsection{Feedback lineage and organization controls}
\label{sec:feedback-controls}

The strongest control is the lineage ablation. The proposal-entropy audit tests whether the stream is repeated sampling, and the paired Parameter Golf memory ablation removes shared lineage while keeping the same starting recipe, specialist split, submitted-trial budget, and current-best score line. Figure~\ref{fig:pg-ablation} summarizes the Parameter Golf controls. Panel A compares the role-decomposed swarm with two agent baselines. The role-decomposed lineage run reaches 1.073142 with 16 valid drops by 200 trials, while the 10-agent generic control reaches 1.074495 with 10 drops. The same-harness single-generalist control finds 14 drops and reaches 1.075384, but its stream is more concentrated: the largest cluster consumes 35 of 200 submissions, including 32 preflight crashes around polar-coefficient edits, versus 7 of 200 for the role swarm. Panel A therefore shows that role-partitioned lineage improves score and boundary discipline under the same budget.

Panel B shows the sharpest feedback effect under a matched 200-trial window. With lineage, the loop finds 16 valid drops and reaches 1.073142 by exp\_176. Without lineage, it finds 3 drops, reaches 1.077413 at exp\_075, then runs 125 submitted trials without a new valid improvement. Lineage acts as active research state by preserving which measured heads remain useful, which edits failed, and which budget boundaries remain. The no-lineage run hits the eval-budget cap on $61.5\%$ of trials versus $19.0\%$ with lineage. The current-best stack already sits close to the $600$\,s eval cap, and only the lineage prompt's Recent Activity block carries that dynamic SOTA fact across sessions. The no-lineage tree collapses to 3 active parent heads versus 15 with lineage, and specialists contributing at least one keep fall from 8 of 10 to 2 of 10.

The generic multi-agent control has 13 declared parent heads, so lineage still maintains multiple measured frontiers, but it hits the eval-budget cap on $59.0\%$ of trials and has only 41.1 effective clusters. The control separates two mechanisms: shared lineage recovers much of the improvement count and parent-tree breadth, while role partitioning improves boundary discipline and broadens the idea stream. Diversity is not required for any valid drops, but here the less diverse stream finds fewer drops, ends at worse bpb, and collides with the eval-budget boundary more often.

\section{Discussion}
\label{sec:discussion}

The final recipes make the loop's scope concrete. They show what the agents developed, how feedback became the next edit, and where current agents stop. The boxes summarize each final approach against its starting recipe, with the trial sequence that produced it. Appendix~\ref{app:final-solutions} gives long-form pipeline descriptions with post hoc explanatory schematics, and Appendix~\ref{app:additional-cases} records lower-level failure and measurement-audit cases.

The boxes use a few recipe-specific terms. Evaluation-time adaptation and TTT-only z-loss are legal score-first Parameter Golf updates. Separate RoPE/NoPE query gains and attention-output gates are compact attention changes. The \texttt{SSSL} to \texttt{L} rewrite and logit bias are NanoChat runtime and model edits. CIFAR warmup repair restores accuracy after shortening the run. Appendix~\ref{app:recipe-glossary} gives definitions.

\begin{tcolorbox}[
  enhanced,
  colback=blue!1,
  colframe=blue!45!black,
  colbacktitle=blue!7,
  coltitle=black,
  title=\small\textbf{Parameter Golf final recipe},
  fonttitle=\bfseries,
  boxrule=0.45pt,
  arc=1mm,
  left=1.8mm,
  right=1.8mm,
  top=1.2mm,
  bottom=1.2mm,
  titlerule=0pt
]
\scriptsize
\setlength{\parskip}{0.25em}

\begin{minipage}[t]{0.48\linewidth}
\textbf{Final developed approach.}
The final recipe improves \texttt{val\_bpb} from 1.0810 to 1.072210, a 0.81\% reduction. It combines legal score-first evaluation-time adaptation with compact structural changes: recurrent residual scaling, separate RoPE and NoPE query gains, a per-head attention-output gate, GPTQ calibration changes, and a TTT-only z-loss objective.
\end{minipage}
\hfill
\begin{minipage}[t]{0.48\linewidth}
\textbf{Trace to final recipe.}
The key trace converts \texttt{587}, a size-blocked z-loss idea at 1.072431 and 2,056 bytes over cap, into size-valid \texttt{596} at 1.072251 and 15,995,930 bytes by recovering artifact headroom. \texttt{746}'s eval-side drop reaches 1.072246 and \texttt{750}'s optimizer refinement reaches 1.072210. The loop turns boundary failure into a valid direction.
\end{minipage}
\end{tcolorbox}

The Parameter Golf case converts a score-useful but artifact-ineligible idea into a valid one once feedback names the byte boundary. The size-blocked z-loss trial returned a measured score and exact byte excess, which led to a follow-up edit that recovered artifact headroom while keeping the same score-first evaluation-time objective.

\begin{tcolorbox}[
  enhanced,
  colback=teal!1,
  colframe=teal!45!black,
  colbacktitle=teal!7,
  coltitle=black,
  title=\small\textbf{NanoChat-D12 final recipe},
  fonttitle=\bfseries,
  boxrule=0.45pt,
  arc=1mm,
  left=1.8mm,
  right=1.8mm,
  top=1.2mm,
  bottom=1.2mm,
  titlerule=0pt
]
\scriptsize
\setlength{\parskip}{0.25em}

\begin{minipage}[t]{0.48\linewidth}
\textbf{Final developed approach.}
The final recipe improves CORE from 0.1618 to 0.2244, a 38.7\% gain. This systems-aware pretraining recipe rewrites the attention path from \texttt{SSSL} to \texttt{L}, moves all 12 layers onto Flash SDPA, spends recovered wallclock on a longer run with target data ratio 12 to 100 to 130, and adds a zero-initialized logit-bias path after \texttt{lm\_head}.
\end{minipage}
\hfill
\begin{minipage}[t]{0.48\linewidth}
\textbf{Trace to final recipe.}
The trace runs from \texttt{007}'s runtime diagnosis, changing \texttt{SSSL} to \texttt{L} and moving all 12 layers onto Flash SDPA, to \texttt{020}'s larger token budget and biggest NC jump, +0.0334 CORE. \texttt{024} raises the data ratio, \texttt{025} lands the 0.2241 schedule plateau, and \texttt{156} adds the zero-initialized logit-bias path to reach 0.2244 CORE. A measured systems fact becomes the next pretraining idea.
\end{minipage}
\end{tcolorbox}

The NanoChat-D12 case shows more than final-score improvement. The attention-path rewrite returned recovered wallclock through lineage as usable headroom. Later proposals spent it on more tokens, then refined the same head with a smaller logit-bias path.

\begin{tcolorbox}[
  enhanced,
  colback=green!1,
  colframe=green!45!black,
  colbacktitle=green!7,
  coltitle=black,
  title=\small\textbf{CIFAR-10 Airbench96 final recipe},
  fonttitle=\bfseries,
  boxrule=0.45pt,
  arc=1mm,
  left=1.8mm,
  right=1.8mm,
  top=1.2mm,
  bottom=1.2mm,
  titlerule=0pt
]
\scriptsize
\setlength{\parskip}{0.25em}

\begin{minipage}[t]{0.48\linewidth}
\textbf{Final developed approach.}
The final recipe reduces training time from 26.3560 s to 25.1464 s while satisfying the 0.96 accuracy gate, a 4.59\% speedup. It skips most logging-only validation calls, shortens the horizon, increases learning-rate intensity, and repairs the accuracy margin with faster warmup.
\end{minipage}
\hfill
\begin{minipage}[t]{0.48\linewidth}
\textbf{Trace to final recipe.}
The trace moves from \texttt{007}'s validation-overhead removal to \texttt{008} and \texttt{030}'s shorter schedules, then uses \texttt{060}'s 25.1650 s near-miss to motivate \texttt{070}'s warmup repair. The evaluator rejects fast but inaccurate code, and the loop turns the near-miss into a valid speed recipe.
\end{minipage}
\end{tcolorbox}

The CIFAR case uses an accuracy-gate miss as a research signal rather than a discard. The evaluator rejects fast but inaccurate code, the near miss returns a measured accuracy deficit, and the next successful edit repairs it while preserving most speed gain.

Together, the cases show auto research as a closed empirical trajectory rather than a one-shot generated artifact. Across headline runs, traces record proposals, executable edits, evaluator outcomes, and follow-up ideas. Submitted ideas go beyond scalar tuning: agents modify attention paths, optimizer updates, loss functions, recurrence scaling, quantization, proxy training, and gate-aware speed recipes, with proposals including GQA K/V projection rewrites, Bigram Hash Embeddings, MTP-2 objectives, self-paced loss caching, residual-preserved ConvGroup depth changes, and the NanoChat \texttt{SSSL} to \texttt{L} attention-path rewrite. Role partitioning assigns priors to recipe surfaces, while matched controls in Section~\ref{sec:experiments} show shared lineage and role-partitioned search broaden the submitted idea stream.

\paragraph{Scope and limits.}
The cases mark what the closed loop can and cannot do under Section~\ref{sec:methodology}'s conditions. The observed boundary is compositional, where agents combine, transfer, and repair known techniques under external feedback while respecting constraints during a multi-day run. The loop is best suited to settings where failures become trusted, compact feedback within a bounded trial budget, and less suited to questions whose evidence is subjective or not automatically verifiable. The recorded trials do not show paradigm-level architecture invention such as a replacement for the original Transformer. Future agents may cross this boundary, and the same evaluator-driven feedback loop remains the natural arena for measuring whether such ideas hold up under real training. Within these limits the loop turns auto research from a one-shot claim into a continuously measured object.

\paragraph{Future work.}
Future work follows the same conditions. Other compute-budgeted environments such as image, speech, or reinforcement-learning recipes can use the loop when trials are affordable and externally verified. Longer runs over weeks of continuous GPU time may expose cross-role composition beyond the matched 200-trial windows, and Eq.~\ref{eq:parallel-efficiency} helps plan that compute. Stronger future agents may also propose paradigm-level ideas rather than only compositional ones; the same evaluator-driven loop can measure whether they hold up under real training data. Releasing traces further enables retrospective human-versus-agent comparisons on the same environments.

\section{Conclusion}
\label{sec:conclusion}

We studied auto research as a closed empirical loop that turns ML research into an inspectable sequence of executable proposals, code edits, evaluator-owned measurements, failures, and follow-up ideas. After one-time setup and launch, specialist agents ran 1,197 headline-run trials plus 600 Parameter Golf control trials by writing code, submitting experiments, reading external feedback, and propagating measured facts through shared lineage without humans choosing proposals or repairing failures during search. Across Parameter Golf, NanoChat-D12, and CIFAR-10 Airbench96, the headline runs improved fixed compute-budgeted recipes by $0.81\%$, $38.7\%$, and $4.59\%$ relative to their starting points. These gains show externally verified progress on real training pipelines rather than plans, reports, or scalar sweeps. Parameter Golf controls identify lineage feedback as a key mechanism for turning measured outcomes into later program-level edits, while NanoChat-D12 and CIFAR-10 case traces show the same pattern under different constraints. In NanoChat-D12, the loop converted a systems fact into more training tokens and a small logit-bias refinement. In Parameter Golf, it turned a score-useful but artifact-ineligible z-loss result into an artifact-valid keep. In CIFAR-10 Airbench96, a measured gate miss led directly to the final warmup repair. Each move is the same feedback loop applied to a different environment.

The main lesson is that closed-loop auto research is useful and measurable when the environment owns the metric, per-trial cost is bounded, and outcomes return quickly enough to affect later proposals. Specialist agents cover many recipe surfaces, and Equation~\eqref{eq:parallel-efficiency} captures how parallel submission scales the loop in continuous time. Within what current language models can compose, the pattern is practical and auditable because shared lineage preserves successes and boundary failures, experiments produce real feedback, and an evaluator the recipe cannot rewrite decides which proposals count. This changes agentic ML research from a final answer into a reusable record of what was tried, why, what failed, what improved, and how the next proposal changed. The same feedback loop can make empirical research more scalable, inspectable, and powerful as models become more capable.

\newpage
\bibliographystyle{plainnat}
\bibliography{reference}

\newpage
\appendix

\section*{Contents of Appendix}
\startcontents[appendix]
\printcontents[appendix]{}{1}[2]{}
\clearpage

\section{Specialist prompt templates}
\label{app:prompts}

Each specialist's session begins with a system prompt assembled from three pieces, in this order:

\begin{enumerate}\itemsep0pt
  \item \textbf{Knowledge files.} Static markdown documents under the task package's \texttt{knowledge/} directory, concatenated and pinned at the top of the system prompt so the Anthropic prompt cache can amortise them across sessions. The set is task-specific and fixed before the reported run starts.
  \item \textbf{Global rules.} A task-level protocol shared by every specialist on that task. Defines hard limits, the tool protocol, and the per-session workflow. Figure~\ref{app:fig:global-rules} reproduces the abridged Parameter Golf version.
  \item \textbf{Domain preamble.} A specialist-specific scope and edit-radius statement. Figures~\ref{app:fig:arch-preamble}, \ref{app:fig:opt-preamble}, and \ref{app:fig:meta-preamble} show three of the ten Parameter Golf preambles; the remaining seven and the NanoChat-D12 / CIFAR preambles follow the same structure (scope + non-scope + edit-radius guidance).
\end{enumerate}

The per-iteration \textbf{user message} is rendered fresh from the live blackboard at every session start. Figure~\ref{app:fig:user-msg-full} shows the full-lineage form. Figure~\ref{app:fig:user-msg-blank} shows the no-lineage ablation form (Section~\ref{app:no-lineage}).

\subsection{Global rules (Parameter Golf, abridged)}
\label{app:prompts:global}

\begin{figure}[t]
\centering
\begin{NewBoxFloat}{Parameter Golf global rules (abridged).}{app:box:global-rules}
{\scriptsize\ttfamily\raggedright
You are one specialist in a multi-agent auto-research swarm working on the Parameter Golf challenge. Your goal every session is to propose a concrete edit to \texttt{train\_gpt.py}, validate it locally (syntax + size), submit via the \texttt{submit\_trial} tool, and learn from the returned row. One submit is a complete session; a second submit is allowed only when the first row surfaces a clear, concrete next edit --- otherwise stop.\par\smallskip
\textbf{Hard limits (enforced by the harness)}\par
-- Submission size $\leq$ 16{,}000{,}000 bytes (code + packed model). Comments and docstrings are auto-stripped before the LZMA pack.\par
-- Train wall $\leq$ 600\,s; eval wall $\leq$ 600\,s.\par
-- Each call to \texttt{submit\_trial} produces one TSV row. Multiple submits per session are allowed; each is independently recorded.\par\smallskip
\textbf{Tool protocol}\par
-- \texttt{cwd} is your \texttt{workdir\_<domain>/}. \texttt{train\_gpt.py} lives there.\par
-- \texttt{Bash} is OS-sandboxed: reads are unrestricted; writes are confined to \texttt{cwd}. \texttt{Write} is not in your allowed tools.\par
-- \texttt{submit\_trial} is the only GPU-burning tool. Use \texttt{syntax\_check} and \texttt{size\_project} freely before submitting.\par
-- \texttt{WebSearch} / \texttt{WebFetch} are the primary research channel for non-trivial design questions; the local PR library is supplemental cross-reference only.\par
-- Typical edit sequence: \texttt{Read train\_gpt.py} $\rightarrow$ \texttt{Edit(old, new)} $\rightarrow$ \texttt{syntax\_check} $\rightarrow$ \texttt{size\_project} $\rightarrow$ \texttt{submit\_trial}.\par\smallskip
\textbf{On-demand knowledge files}\par
\texttt{../knowledge/LESSONS.md} carries task-local setup notes; read when relevant, otherwise ignore.\par\smallskip
\textbf{Workflow each session}\par
1. Read \texttt{LEADERBOARD}, \texttt{KNOWLEDGE.md}, and Recent Activity in the user message; identify the current best.\par
2. Decide: mutate from the best, or rebase onto a non-best snapshot via \texttt{rebase\_to}.\par
3. Mutate \texttt{train\_gpt.py} via \texttt{Edit}.\par
4. Call \texttt{syntax\_check} and \texttt{size\_project}; fix and retry on failure.\par
5. Call \texttt{submit\_trial} with a one-sentence hypothesis and a signed \texttt{expected\_delta}.\par
6. Reflect on the result. One submit is a complete session; repeat from step 1 only if the returned row points to a specific next edit.\par
}
\end{NewBoxFloat}
\caption{Parameter Golf \texttt{GLOBAL\_RULES}, abridged. The full source is in \texttt{multi\_agent\_pg/agents/prompts.py}. NanoChat-D12 and CIFAR have analogous global rules with task-specific limits (e.g.\ NC's 90-minute pretraining cap, CIFAR's $0.96$ accuracy gate).}
\label{app:fig:global-rules}
\end{figure}

\subsection{Per-domain preambles}
\label{app:prompts:domains}

\begin{figure}[t]
\centering
\begin{NewBoxFloat}{Parameter Golf \textbf{Architecture} domain preamble.}{app:box:arch}
{\scriptsize\ttfamily\raggedright
You are the \textbf{Architecture} specialist. Your scope is the transformer block itself: attention variants (full, sliding, differential, MLA), recurrence modules (GLA/Mamba/RWKV-style SSMs), residual topology (parallel vs sequential, Pre-Norm vs DeepNorm), MLP variants (SwiGLU, GeGLU, gated MoE-lite), normalisation (RMSNorm, sub-LN), embedding schemes (tied, factored, RoPE/ALiBi/xPos). You do NOT own optimizer, loss, dataset, or quantization --- those are other specialists' domains. Small architectural tweaks (layer count $\pm 1$, dim $\pm 64$) are fine, but prefer changes that cross a qualitative line (e.g.\ swap a block type, add/remove a residual) when the current best has already been small-tweaked to death.\par\smallskip
\textbf{Edit radius:} your domain's historical wins come from structural changes --- block type swap, residual topology flip, norm placement, attention head grouping. A single scalar tweak to an existing module (\texttt{init\_std} 0.005$\rightarrow$0.008) is hparam noise, not architecture --- that belongs in \texttt{opt} or \texttt{meta}. If your draft hypothesis is ``change one number'', you're probably in the wrong domain; pivot to a qualitative edit.\par
}
\end{NewBoxFloat}
\caption{One of ten Parameter Golf domain preambles. Each preamble follows the same three-part shape: scope (what the role owns), non-scope (what the role does not touch), and edit radius (the level of intervention that historically produced wins for this role).}
\label{app:fig:arch-preamble}
\end{figure}

\begin{figure}[t]
\centering
\begin{NewBoxFloat}{Parameter Golf \textbf{Optimizer} domain preamble.}{app:box:opt}
{\scriptsize\ttfamily\raggedright
You are the \textbf{Optimizer} specialist. Scope: optimizer algorithm (Muon variants, Lion, Shampoo, Sophia), learning-rate schedule (cosine, WSD, linear warmup, per-param decay), momentum and weight-decay coupling, gradient clipping, LAWA/EMA weight averaging, per-tensor LR scaling. You do NOT edit model architecture or the loss function. Most of the value here lives in matching schedule shape to the 600\,s budget --- do not propose schedules that implicitly assume more or fewer steps than the current best trains for.\par\smallskip
\textbf{Edit radius:} your domain's wins come from schedule-shape changes --- swap the schedule family (cosine $\rightarrow$ WSD, linear $\rightarrow$ triangle), introduce a new warmdown phase, apply a different optimizer family to one parameter group, couple momentum$\leftrightarrow$LR in a new way. Single-coefficient tweaks (\texttt{muon\_wd} 0.095$\rightarrow$0.110, \texttt{adam\_eps} 1e-8$\rightarrow$1e-9) rarely exceed Fisher-info noise at our $\Delta$ scale unless they cross a qualitative threshold.\par
}
\end{NewBoxFloat}
\caption{The Parameter Golf \textbf{Optimizer} preamble. Compared with \textbf{Architecture}, the scope is narrower (schedules, weight decay, optimizer family) but the same scope/non-scope/edit-radius shape is preserved.}
\label{app:fig:opt-preamble}
\end{figure}

\begin{figure}[t]
\centering
\begin{NewBoxFloat}{Parameter Golf \textbf{Meta-Search} (analyst) domain preamble.}{app:box:meta}
{\scriptsize\ttfamily\raggedright
You are the \textbf{Meta-Search} analyst. Scope: hyperparameter sweeps across the recent kept trials --- LR multipliers, batch size, warmup ratio, weight-decay, init scale. You are an ANALYST: you mostly read \texttt{results.tsv} + \texttt{KNOWLEDGE.md} to find a narrow hyperparameter tweak that several prior trials missed. Keep each tweak small-radius --- hyperparameter moves, not structural changes.\par\smallskip
\textbf{Edit radius:} your domain IS ``small-radius'', but the radius should be on axes with HIGH unexplored volume, not crowded knobs. Before proposing, slice \texttt{results.tsv} via \texttt{Bash} to check how many trials already touched your proposed knob --- if $> 5$ in your own domain's recent window, that axis is crowded; find a different one. For fresh axes, \texttt{WebSearch} recent hyperparameter-tuning literature first; the PR library's \texttt{gaps.md} is a secondary reference for what has been ruled out in the environment.\par
}
\end{NewBoxFloat}
\caption{The \textbf{Meta-Search} preamble explicitly frames the role as an analyst who reads existing results before proposing. This is the only role in Parameter Golf that is allowed to lean on the lineage as its primary input rather than as a check on a freshly proposed edit.}
\label{app:fig:meta-preamble}
\end{figure}

The remaining seven Parameter Golf preambles (\texttt{quant}, \texttt{tok}, \texttt{ttt}, \texttt{curr}, \texttt{loss}, \texttt{reg}, \texttt{eval}) follow the same shape and are reproduced verbatim in the source repository at \texttt{multi\_agent\_pg/agents/prompts.py}. NanoChat-D12 has five preambles (\texttt{arch}, \texttt{opt}, \texttt{data}, \texttt{sched}, \texttt{sys}) and CIFAR-10 Airbench96 has five (\texttt{arch}, \texttt{opt}, \texttt{aug}, \texttt{loss}, \texttt{reg}). Per-task preambles never refer to other tasks' constraint regimes.

\subsection{Anti-anchoring and crash feedback}
\label{app:prompts:anti-anchor}

Specialists can repeatedly return to high-salience edits, which collapses the proposal stream toward a small set of canonical moves. Each specialist receives a short banlist of patterns that earlier sessions in its own role have already tried and that failed or returned within noise. The banlist is rendered alongside the lineage slice so the same dead end is less likely to be re-explored.

Crash handling uses the same prompt channel. When a submitted trial crashes, the lineage slice for the next trial carries the deepest exception line and the deepest training-script frame from the crash. This gives the next specialist an explicit failure mode rather than a generic crash status, which makes surface variants of the same failed proposal less likely.

\subsection{Per-iteration user message}
\label{app:prompts:user-msg}

\begin{figure}[t]
\centering
\begin{NewBoxFloat}{Per-iteration user message --- full lineage.}{app:box:user-msg-full}
{\scriptsize\ttfamily\raggedright
\# Session start --- 2026-04-22T01:36:09Z\par
You are specialist \textbf{arch}. Your workdir is \texttt{<workdir\_arch>}.\par
Current best: \textbf{exp\_037} (val\_bpb=1.07683).\par\smallskip
\#\# LEADERBOARD.md\par
\textit{<top-N keep rows by score, in compact bullet form>}\par\smallskip
\#\# KNOWLEDGE.md\par
\textit{<curated tree of prior hypotheses + outcomes; agent-readable summary>}\par\smallskip
\#\# Recent Activity (most recent 10)\par
- exp\_046 [arch, discard, val\_bpb=1.0775, $\Delta$=+0.0007] differential attention with cross-thread V residual\par
- exp\_045 [opt,  keep,    val\_bpb=1.07650, $\Delta$=$-$0.0003] Muon WSD warmdown 25\%$\rightarrow$30\%\par
- ...\par\smallskip
\#\# (optional) $\Delta$ Saturation signal\par
\textit{<emitted only when last 5 own-role trials all $|\Delta|<5\times10^{-4}$>}\par\smallskip
\#\# Your workdir\par
- \texttt{train\_gpt.py} present, 41{,}823 bytes\par
\hphantom{XX}head: \texttt{"""train\_gpt.py --- 1.0810 SOTA stack..."""}\par\smallskip
\#\# Your task this session\par
Propose an edit within the \textbf{arch} domain, validate it locally, and submit via \texttt{submit\_trial}. A single submit is a complete session; a second submit is allowed only when the first row points to a concrete next edit.\par
}
\end{NewBoxFloat}
\caption{Schematic of the user message rendered for each specialist session under full lineage. Concrete content is drawn from \texttt{LEADERBOARD.md}, \texttt{KNOWLEDGE.md}, and the most-recent ten rows of \texttt{results.tsv}. The exact assembly is in \texttt{multi\_agent\_core/agents/base.py:render\_user\_message}.}
\label{app:fig:user-msg-full}
\end{figure}

\begin{figure}[t]
\centering
\begin{NewBoxFloat}{Per-iteration user message --- no-lineage ablation.}{app:box:user-msg-blank}
{\scriptsize\ttfamily\raggedright
\# Session start --- 2026-05-01T04:21:37Z\par
You are specialist \textbf{arch}. Your workdir is \texttt{<workdir\_arch>}.\par
Current best: \textbf{exp\_000} (val\_bpb=1.081).\par\smallskip
\#\# Your workdir\par
- \texttt{train\_gpt.py} present, 41{,}823 bytes\par\smallskip
\#\# Your task this session\par
Propose an edit within the \textbf{arch} domain, validate it locally, and submit via \texttt{submit\_trial}. A single submit is a complete session; a second submit is allowed only when the returned row points to a concrete next edit.\par\smallskip
\#\# Lineage policy (this run)\par
Within-run prior-trial logs are unavailable for this run. The \texttt{LEADERBOARD} / \texttt{KNOWLEDGE} / Recent Activity sections you may have seen in other runs are intentionally absent. The \texttt{read\_snapshot} and \texttt{diff\_snapshots} tools are also disabled. Do NOT attempt to query blackboard files via \texttt{Bash} (\texttt{tree.tsv}, \texttt{results.tsv}, \texttt{lineage\_snapshots/}, \texttt{events.jsonl}, \texttt{best.json}, \texttt{supervisor\_audit.jsonl}, anything under \texttt{blackboard/}) --- those reads are rejected at the harness level. Propose from your static setup priors in the system prompt, the current-best score above, your workdir state, and your own in-session reasoning.\par
}
\end{NewBoxFloat}
\caption{Schematic of the user message under the no-lineage ablation (\texttt{MAGENT\_NO\_LINEAGE=1}). Sections that read within-run prior-trial outcomes are omitted. The current-best \texttt{exp\_id} and score are preserved because the agent uses them to root \texttt{rebase\_to(best, workdir)} at session start. See Section~\ref{app:no-lineage} for the operational definition.}
\label{app:fig:user-msg-blank}
\end{figure}

\section{Tool catalogue}
\label{app:tools}

Each agent session is wired with a small set of MCP tools (Table~\ref{app:tab:tools}) plus the SDK built-ins \texttt{Read} / \texttt{Edit} / \texttt{Bash} / \texttt{Grep} / \texttt{Glob} / \texttt{WebSearch} / \texttt{WebFetch} / \texttt{ToolSearch} / \texttt{TodoWrite} / \texttt{Agent}. \texttt{Write} is deliberately not in the allowed-tools list: the artifact packer ignores any file that is not the canonical recipe file, so sidecars created via \texttt{Write} would silently inflate the artifact toward the size cap.

Specialists can use web search and task-local knowledge files. The local library is helpful for known moves. Web search is most useful when a task is newly forked or when a failure points to a runtime dependency. The two channels provide complementary context for proposing and diagnosing submitted trials.

\begin{table}[t]
\centering
\footnotesize
\caption{In-process MCP tools surfaced to specialists. Names are namespaced \texttt{mcp\_\_apg\_\_<name>} on the SDK tool channel; the model sees both name and JSON-schema description automatically. Source: \texttt{multi\_agent\_core/tools/} and \texttt{multi\_agent\_pg/tools/}.}
\label{app:tab:tools}
\setlength{\tabcolsep}{4.5pt}
\renewcommand{\arraystretch}{1.16}
\begin{tabularx}{\textwidth}{@{}
>{\raggedright\arraybackslash\ttfamily}p{0.16\textwidth}
>{\raggedright\arraybackslash}X
>{\raggedright\arraybackslash}p{0.27\textwidth}
@{}}
\toprule
\rmfamily Tool & Description (one-liner) & Required arguments \\
\midrule
submit\_trial &
Submit the specialist's current \texttt{train\_gpt.py} to a real eight-H100 evaluation. Runs local syntax + size preflight first; failures are recorded without GPU time. Blocks until the job finishes, then writes a row to \texttt{results.tsv}. &
\texttt{specialist}, \texttt{hypothesis}, \texttt{expected\_delta}, \texttt{parent\_exp} \\
\midrule
syntax\_check &
\texttt{py\_compile} the editable file and report any \texttt{SyntaxError} without executing. Millisecond-scale; catches the most common edit mistake before a GPU trial. &
\texttt{workdir} \\
\midrule
size\_project &
Run the real \texttt{lzma+base85} pack step locally and report the projected packed size against the 16\,MB cap. Used before \texttt{submit\_trial} to catch oversize edits without burning a job. &
\texttt{workdir} \\
\midrule
param\_count &
Static AST estimate of trainable parameter count: sums \texttt{nn.Linear} / \texttt{nn.Embedding} literal sizes. Fast ($\sim$5\,ms) but only catches gross structural changes. &
\texttt{workdir} \\
\midrule
read\_snapshot &
Fetch the snapshotted source of a past kept experiment to study what a sibling specialist actually wrote. Truncates to $\sim$200\,KB. Disabled under \texttt{MAGENT\_NO\_LINEAGE=1}. &
\texttt{exp\_id} (optional \texttt{path}) \\
\midrule
rebase\_to &
Copy a past experiment's snapshotted source into the current workdir, overwriting whatever is there. Used to fork from a non-best parent. &
\texttt{exp\_id}, \texttt{workdir} \\
\midrule
diff\_snapshots &
Unified diff of two past experiments' snapshotted source. Truncates to $\sim$300 lines / 8\,KB. Disabled under \texttt{MAGENT\_NO\_LINEAGE=1}. &
\texttt{exp\_a}, \texttt{exp\_b} \\
\midrule
read\_pr\_library &
Fetch entry \texttt{N} from the curated PR library: technique, specialist tag, risk tag, available file paths. Cross-reference for web-found ideas; not the primary research source. &
\texttt{pr\_number} \\
\midrule
read\_pr\_source &
Return the extracted source text from one file inside a PR library entry. &
\texttt{pr\_number}, \texttt{path} \\
\bottomrule
\end{tabularx}
\end{table}

Three SDK hooks moderate the tool channel:
\begin{itemize}\itemsep0pt
  \item \textbf{\texttt{block\_bash\_writes}} (PreToolUse, on \texttt{Bash}): denies destructive shell verbs (\texttt{rm}, \texttt{mv}, \texttt{cp}, in-place \texttt{sed}, \texttt{tar -c}, file redirects, package install, process control). Bash is read-only in the swarm.
  \item \textbf{\texttt{block\_bash\_blackboard}} (PreToolUse, on \texttt{Bash}): denies reads of blackboard files (\texttt{tree.tsv}, \texttt{results.tsv}, \texttt{lineage\_snapshots/}, \texttt{events.jsonl}, \texttt{best.json}, \texttt{supervisor\_audit.jsonl}, anything under \texttt{blackboard/}) when \texttt{MAGENT\_NO\_LINEAGE=1}. Pass-through otherwise.
  \item \textbf{\texttt{cap\_builtin\_tool\_output}} (PostToolUse, on \texttt{Bash} / \texttt{Grep} / \texttt{WebFetch} / \texttt{WebSearch}): truncates oversized outputs at 16\,KB with a recovery marker, bounding cache growth without breaking legitimate slicing.
\end{itemize}

\section{Trial classification}
\label{app:status}

The harness assigns one of nine status values to every submitted trial. Classification is performed by a post-trial parser (\texttt{multi\_agent\_<task>/tools/run\_classify.py}) that reads the combined preflight + train + pack log and emits a JSONL row. The blackboard then maps that coarse status to the per-task enum in Table~\ref{app:tab:status}.

\begin{table}[t]
\centering
\footnotesize
\caption{Trial status taxonomy. Source rules: \texttt{multi\_agent\_pg/tools/run\_classify.py}, \texttt{multi\_agent\_core/harness/tracker.py}, and per-task adapters.}
\label{app:tab:status}
\setlength{\tabcolsep}{5pt}
\renewcommand{\arraystretch}{1.16}
\begin{tabularx}{\linewidth}{@{}
>{\raggedright\arraybackslash\ttfamily}p{0.27\linewidth}
>{\raggedright\arraybackslash}X
@{}}
\toprule
\rmfamily Status & Triggering condition \\
\midrule
baseline                  & The single seed row written by \texttt{bootstrap\_from\_baseline}; not a submitted agent trial. \\
\midrule
keep                      & Valid measured run AND strictly better than the prior best in the metric's preferred direction. \\
\midrule
discard                   & Valid measured run AND not strictly better than the prior best. \\
\midrule
crash                     & Train phase exited non-zero AND no valid score parse, OR pack step failed but train ran. \\
\midrule
preflight\_crash          & Local syntax check failed OR GPU-cluster submission failed before any GPU time was burned. \\
\midrule
size\_blocked             & Submission size $>$ 16\,MB at preflight (smoke pack) or post-run pack. \\
\midrule
train\_budget\_overrun    & Train phase exceeded the task train budget plus 5\,s tolerance for step-atomic granularity. \\
\midrule
eval\_budget\_overrun     & Eval phase exceeded 600\,s, or the outer trial timeout fired before eval completed. No tolerance: eval is continuous wallclock. \\
\midrule
disqualified (CIFAR only) & Mean accuracy across cold-process seeds fell below the strict $0.96$ gate. \texttt{train\_s} is blanked and the row cannot win regardless of speed. \\
\midrule
harness\_abort            & Bookkeeping-side failure (e.g.\ cluster scheduler bookkeeping loss); not a substantive signal. Quarantined from the prompt's Recent Activity. \\
\bottomrule
\end{tabularx}
\end{table}

\section{Run configuration}
\label{app:config}

\subsection{Per-task swarm configuration}

Each task package ships a \texttt{swarm\_config.json} that is the single source of truth for two per-specialist knobs: model assignment and GPU-cluster priority. The supervisor logs the resolved values at startup (\texttt{supervisor\_audit.jsonl}).

\begin{table}[t]
\centering
\footnotesize
\caption{Resolved per-specialist model assignment and GPU-cluster priority for each task.}
\label{app:tab:swarm-config}
\setlength{\tabcolsep}{5pt}
\renewcommand{\arraystretch}{1.16}
\begin{tabularx}{\linewidth}{@{}
>{\raggedright\arraybackslash}p{0.18\linewidth}
>{\raggedright\arraybackslash}X
>{\raggedright\arraybackslash}p{0.18\linewidth}
@{}}
\toprule
Task & Model assignment & GPU-cluster priority \\
\midrule
Parameter Golf &
All ten roles on Claude Opus 4.7. &
All ten roles at priority 10. \\
\midrule
NanoChat-D12 &
All five roles on Claude Opus 4.7. &
\texttt{arch} / \texttt{opt} / \texttt{data} at 10; \texttt{sched} / \texttt{sys} at 9. \\
\midrule
CIFAR-10 Airbench96 &
All five on Claude Opus 4.7. &
\texttt{arch} / \texttt{opt} / \texttt{aug} at 10; \texttt{loss} / \texttt{reg} at 9. \\
\bottomrule
\end{tabularx}
\end{table}

\subsection{Model routing}

The framework supports per-role model routing, because different roles may need different reasoning depth. Architecture proposals tend to require deeper combinatorial search, while optimization, augmentation, and schedule proposals tend to be more tactical. All reported frozen runs use Claude Opus 4.7 for every role; only GPU-cluster priority differs by role. The routing policy is declared in \texttt{swarm\_config.json} and kept fixed during each frozen run.

\subsection{Doer session defaults}

A specialist session is a single \texttt{ClaudeSDKClient} call. Defaults from \texttt{multi\_agent\_core/agents/base.py:DoerConfig}:

\begin{itemize}\itemsep0pt
  \item \texttt{thinking\_budget = 8000} tokens of extended thinking budget.
  \item \texttt{max\_turns = 200} per session (cap on tool-use turns; sized for multi-submit sessions with PR-library drill-down).
  \item \texttt{enable\_web = True} (\texttt{WebSearch} / \texttt{WebFetch} are forced into the SDK's preload list because the default preset would leave them deferred).
  \item Sandbox: \texttt{bubblewrap}-based with \texttt{Bash} read-anywhere / write-only-to-cwd. On hosts where pivot-root is unavailable (LXC, nested containers), the sandbox is auto-disabled and the \texttt{block\_bash\_writes} hook becomes the primary write barrier.
  \item Permission mode: \texttt{bypassPermissions} (autonomous; the allowed-tools list is the safety boundary, not interactive prompting).
\end{itemize}

\subsection{Termination rules}

The supervisor stops on the OR of two conditions (\texttt{multi\_agent\_core/supervisor/termination.py}):

\begin{itemize}\itemsep0pt
  \item Wall-clock deadline: default 48\,h, configurable via \texttt{--deadline-hours}.
  \item No-improvement grace: default 4\,h since the most recent \texttt{keep}, configurable via \texttt{--no-improvement-hours}. The grace clock resets every time a new \texttt{keep} lands.
\end{itemize}

A \texttt{stop.flag} file is written under \texttt{blackboard/} when either condition fires; in-flight specialist coroutines exit at their next \texttt{should\_stop()} check, with a 60\,s grace before forced cancellation. SIGINT and SIGTERM trigger the same shutdown chain plus a best-effort stop sweep over registered GPU-cluster jobs.

\section{Hardware and execution environment}
\label{app:hardware}

All trials run on an internal GPU cluster. Each Parameter Golf and NanoChat-D12 trial is a fresh eight-H100 worker; CIFAR-10 Airbench96 trials run inside a long-lived GPU worker, which preserves the pre-warmed CUDA context across cold-process seeds. Supervisor and dashboard processes run on a head node with local ext4 storage. The blackboard, workdirs, and event logs all live on head-node local storage; only the editable workdir is synchronized to the worker-visible shared filesystem at trial submission time, and only the trial's stdout and artifact are synchronized back. The Anthropic SDK uses the bundled \texttt{claude} CLI binary, which performs HTTP-level retries on transient errors (rate limits, network) before any exception is surfaced to the supervisor's session-level retry path.

\paragraph{Compute accounting.}
The reported submitted-trial counts include valid improvements, valid non-improvements, and failed submitted trials, so the headline and control totals already include most failed-run compute inside the frozen search loops. A conservative cap-derived upper bound for active accelerator time is 4{,}000 H100-hours for the 1{,}500 reported Parameter Golf submitted trials (900 headline plus 600 controls, eight H100s, at most 600\,s train plus 600\,s eval per trial) and 2{,}400 H100-hours for the 200 NanoChat-D12 headline trials (eight H100s, at most 90 minutes per trial). CIFAR-10 Airbench96 is much smaller: 97 submitted trials run on one long-lived GPU worker, and even counting ten cold-process seeds per trial at roughly the reported 25--27\,s scale gives under 10 single-GPU-hours of active training. These bounds exclude queue idle time and are upper bounds rather than summed per-job telemetry; preflight failures that terminate before a worker is launched consume less than the cap. The full project used additional setup and preliminary compute outside these reported totals, including benchmark preparation, starting-point calibration, harness and prompt development, and historical audit traces. Those runs are not counted as headline or control trials because they are not part of the frozen reported search loops.

\section{Starting-point calibration}
\label{app:baseline}

The Parameter Golf, NanoChat-D12, and CIFAR-10 Airbench96 starting points used in Table~\ref{tab:main} were established as follows.

\paragraph{Parameter Golf.} The starting point is the public 1.0810 record published on the Parameter Golf leaderboard. The search uses the corresponding starting code path, but the denominator remains the public score rather than a newly calibrated local baseline. The search delta is therefore reported against the public number.

\paragraph{NanoChat-D12.} The starting point ($0.1618$) is one full run of the unmodified upstream NanoChat-D12 recipe at the pinned commit, scored by parsing the CORE line from the training log. This calibrated upstream run is both the baseline and the fixed search starting point for the reported 38.7\% improvement. Calibration is append-only (\texttt{python -m multi\_agent\_nc.calibrate\_baseline --score 0.1618}) so a populated run log is not silently overwritten. Without local calibration, the public ``recipe starting point'' from upstream NanoChat would be a stale denominator because the score depends on hardware, runtime image, and offline-mode settings.

\paragraph{CIFAR-10 Airbench96.} The starting point ($26.3560$\,s) is a ten-seed cold-process aggregate of the unmodified Airbench96 recipe under the strict $0.96$ accuracy gate. The upstream reported reference of $27.3000$\,s is the same upstream recipe measured under a different reporting protocol, not an agent result and not the denominator for the 4.59\% improvement. The cold-process protocol re-imports the recipe in a fresh Python process per seed so transient compile state does not bias the wallclock measurement. The reported search trajectory and selected recipe use the same cold-process protocol, so the timing source and accuracy gate match the main-paper CIFAR result.

\section{Additional trace statistics}
\label{app:trace-stats}

This section reports role-level trace statistics that support the main text but are secondary to the closed-loop trajectory evidence.

\paragraph{Allocation balance.}
Parameter Golf assigned 84 to 96 trials per role with CV 0.049. NanoChat-D12 assigned 33 to 45 trials per role with CV 0.100. CIFAR-10 Airbench96 assigned 18 to 21 trials per role with CV 0.062. The main differences across tasks therefore come from the mix of valid improvements, valid non-improvements, and ineligible outcomes rather than from one role receiving most of the budget.

\begin{figure}[t]
\centering
\includegraphics[width=\textwidth]{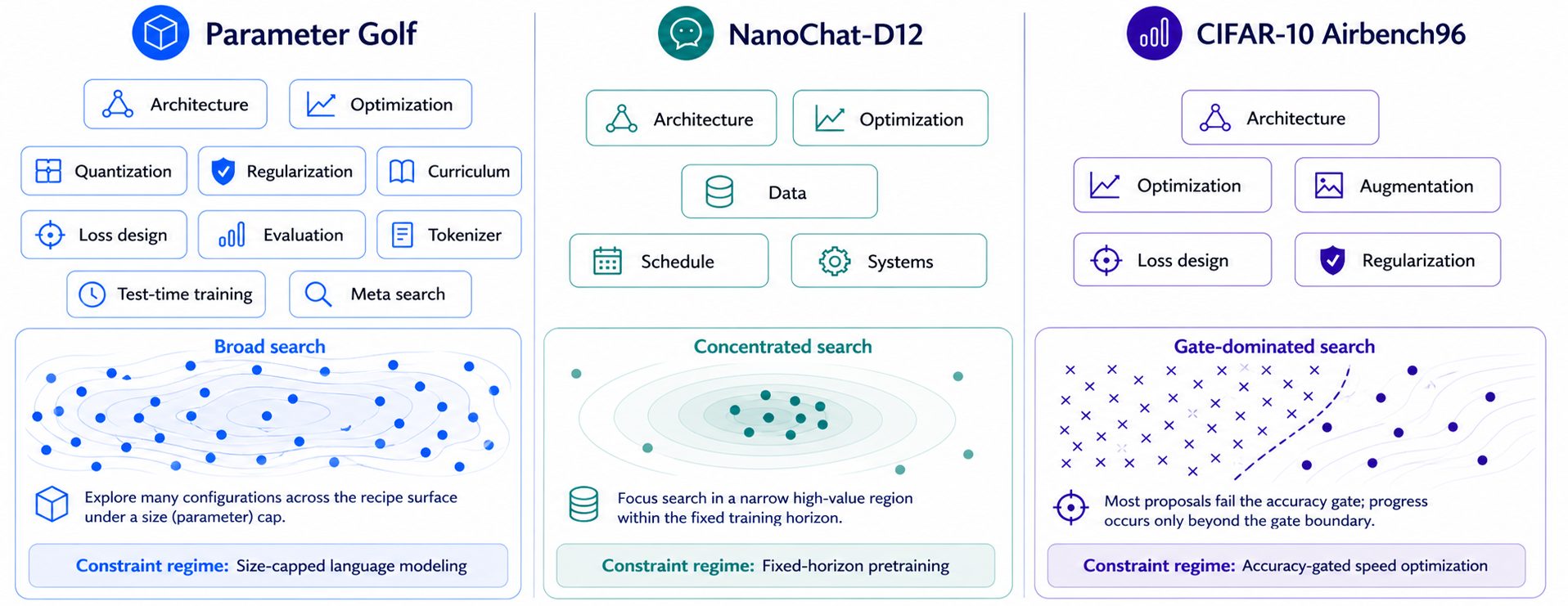}
\caption{Specialist role partitioning and search behavior across environments. The top row shows the fixed role split chosen before search, and the bottom row sketches the resulting search pattern under each constraint regime.}
\label{app:fig:specialist-design}
\end{figure}

\begin{figure}[t]
\centering
\includegraphics[width=\textwidth]{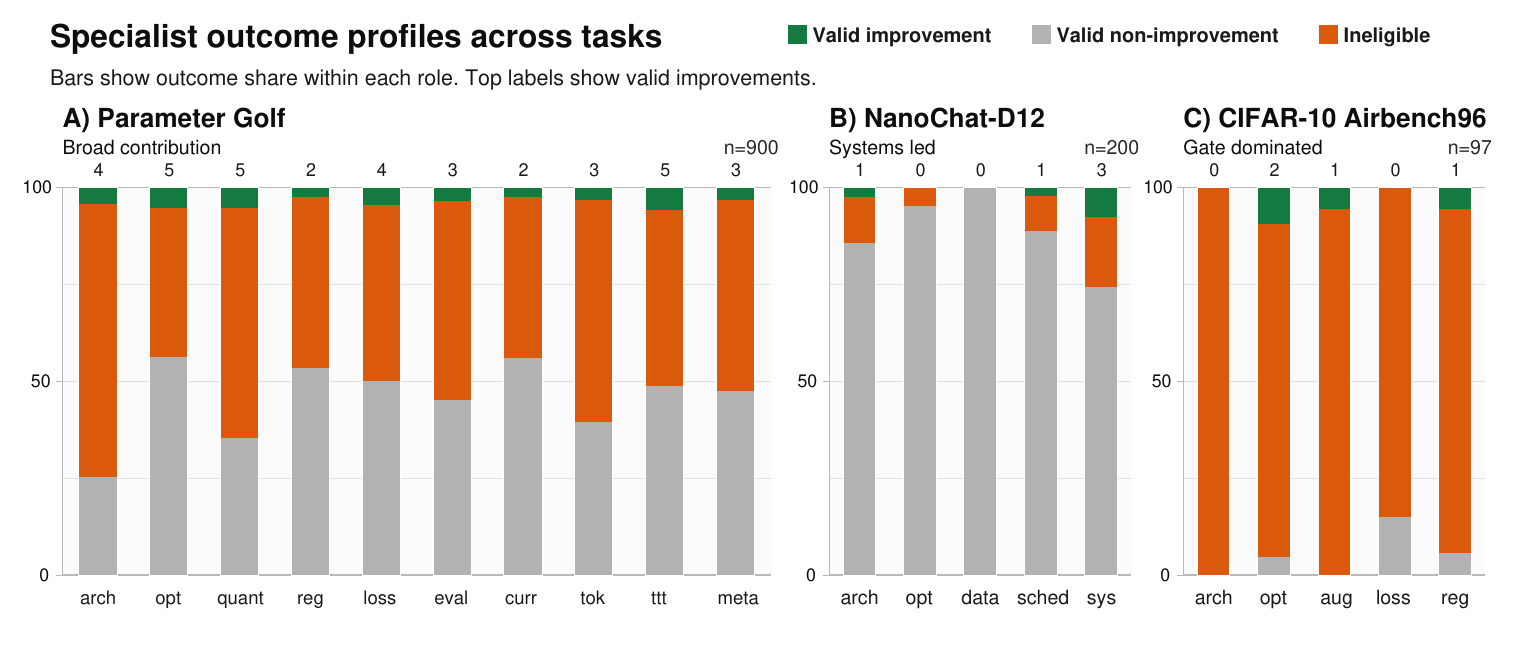}
\caption{Specialist outcome profiles across the three environments. Each stacked bar is normalized within one specialist's submitted trials. Green marks valid improvements, gray marks valid non-improvements, and orange marks ineligible trials. Labels above bars give valid improvement counts.}
\label{app:fig:specialist-outcomes}
\end{figure}

\begin{table}[t]
\centering
\footnotesize
\caption{Specialist contribution patterns by task. Allocation was balanced across roles, so the table reports valid improvements and the dominant valid or ineligible boundary.}
\label{app:tab:specialist-behavior}
\setlength{\tabcolsep}{4.5pt}
\renewcommand{\arraystretch}{1.12}
\begin{tabularx}{\linewidth}{@{}
>{\raggedright\arraybackslash}p{0.19\linewidth}
>{\raggedright\arraybackslash}p{0.15\linewidth}
>{\raggedright\arraybackslash}p{0.33\linewidth}
>{\raggedright\arraybackslash}X
@{}}
\toprule
Task & Search shape & Valid improvements & Dominant boundary \\
\midrule
Parameter Golf &
Broad &
All 10 roles contributed. Opt., TTT, and quant. contributed 5 each. &
Size gate, especially architecture, quantization, and tokenizer. \\
\midrule
NanoChat-D12 &
Concentrated &
Systems produced 3. Schedule and architecture produced 1 each. &
Most non-systems roles produced valid non-improvements. \\
\midrule
CIFAR-10 Airbench96 &
Gate dominated &
Optimization produced 2. Augmentation and regularization produced 1 each. &
81 of 97 trials missed the accuracy gate. \\
\bottomrule
\end{tabularx}
\end{table}

Tool-use patterns provide context but do not explain the task differences by themselves. In Parameter Golf, validation calls were frequent across roles, with 2.39 to 2.96 calls per trial. In NanoChat-D12, optimization used web search most often, but the systems role produced most valid improvements. In CIFAR-10 Airbench96, web and validation use were comparatively uniform. The stronger signal is how each role interacts with the task constraint.

\subsection{Historical single-generalist comparison}
\label{app:single-generalist-caveat}

The historical Parameter Golf single-generalist trace is useful as an audit of submitted proposal diversity, but it is not the primary causal control. Under the common rule that any legal lower \texttt{val\_bpb} counts as a valid improvement, the first 91 single-generalist trials produced 3 valid drops and reached a best reduction of 0.00122 bpb. The original specialist swarm produced 7 valid drops in the same 91-trial window and reached a best reduction of 0.00406 bpb. Four of those 7 swarm drops have declared parents that also fall inside the 91-trial window, which is the denominator used for the historical swarm keep-edge fraction in Table~\ref{tab:proposal-entropy}.

This comparison changes role decomposition, the number of concurrently active proposal threads, and harness vintage. The single-agent run also predates the later anti-anchoring prompt revisions. We therefore use Figure~\ref{fig:pg-ablation} as the primary Parameter Golf control in the main text. The historical trace remains in Table~\ref{tab:proposal-entropy} because it is directly useful for auditing proposal diversity and context partitioning.

\section{No-lineage ablation definition}
\label{app:no-lineage}

The no-lineage ablation closes three lineage feedback channels under a single \texttt{MAGENT\_NO\_LINEAGE=1} environment switch (set via the supervisor's \texttt{--no-lineage} CLI flag).

\begin{enumerate}\itemsep0pt
  \item \textbf{Per-iteration prompt rendering.} The user message rendered at every session start is short-circuited: the \texttt{LEADERBOARD.md} / \texttt{KNOWLEDGE.md} / Recent Activity / Saturation-warning sections are dropped. The current-best \texttt{exp\_id} + score one-liner is preserved (the agent uses it to root \texttt{rebase\_to}). Figure~\ref{app:fig:user-msg-blank} shows the resulting form.
  \item \textbf{Lineage-reading tools.} \texttt{read\_snapshot} and \texttt{diff\_snapshots} are removed from \texttt{allowed\_tools} and \texttt{preload\_tools}. \texttt{rebase\_to} is preserved because it does not return prior-trial \emph{content} to the agent --- it copies code into the workdir using the already-known current-best \texttt{exp\_id}.
  \item \textbf{Bash reads of blackboard files.} The \texttt{block\_bash\_blackboard} \texttt{PreToolUse} hook (Section~\ref{app:tools}) rejects \texttt{Bash} commands matching any of \texttt{tree.tsv}, \texttt{results.tsv}, \texttt{lineage\_snapshots/}, \texttt{events.jsonl}, \texttt{best.json}, \texttt{supervisor\_audit.jsonl}, or any path under \texttt{blackboard/}. This closes the in-practice dominant residual: an empirical audit of the lineage-on Parameter Golf run found that $57.9\%$ of \texttt{Bash} calls with parseable arguments targeted blackboard files (mainly \texttt{awk} slices of \texttt{tree.tsv}); without this hook, the prompt-side ablation would leak heavily.
\end{enumerate}

What is intentionally \emph{not} ablated:

\begin{itemize}\itemsep0pt
  \item \textbf{Static priors.} The system-prompt knowledge files and the global rules + domain preamble are kept. These are task setup priors fixed before the reported run starts; removing them would test ``zero-shot'' agent behaviour, not the value of within-run feedback memory.
  \item \textbf{Current-best score.} The agent sees \texttt{Current best: exp\_<id> (val\_bpb=...)} as a one-line entry. Without this, the agent cannot \texttt{rebase\_to} a usable starting code state and the loop is no longer comparable.
  \item \textbf{Workdir code state.} Each specialist's \texttt{workdir\_<role>/train\_gpt.py} is the agent's edit target and necessarily reflects the current-best code. The agent can \texttt{Read} this file. This is the ablation's irreducible residual: an editable closed-loop process must give the agent something to edit.
\end{itemize}

The ablation therefore tests whether agents can produce diverse, valid proposals when given (a) static priors, (b) the current-best code, and (c) their own in-session reasoning, but NOT (d) any record of within-run prior trials' hypotheses, scores, statuses, diffs, or crash logs.

\section{Final recipe and additional trace details}
\label{app:additional-cases}

The main text presents the final developed approaches at a high level.
This appendix records the supporting trace details that are useful for
audit but too fine-grained for the main case analysis.

\begin{table}[t]
\centering
\footnotesize
\caption{Full representative submitted program transformations. Rows include valid and failed trials because they summarize generated research ideas, not only final-best contributors.}
\label{app:tab:program-transforms-full}
\setlength{\tabcolsep}{4pt}
\renewcommand{\arraystretch}{1.1}
\begin{tabularx}{\textwidth}{@{}
    >{\raggedright\arraybackslash}p{0.15\textwidth}
    >{\raggedright\arraybackslash}p{0.16\textwidth}
    >{\raggedright\arraybackslash}X
@{}}
\toprule
Environment & Trial id(s) & Concrete architecture or program change \\
\midrule

\multirow{2}{0.15\textwidth}{\raisebox{-0.35\baselineskip}{Parameter Golf}}
& \texttt{001, 030/188}
& Value-residual attention thread; parameter-neutral SwiGLU MLP replacing the non-gated squared activation. \\
\cmidrule(l){2-3}
& \texttt{245, 475, 538}
& Recurrent residual scaling; separate RoPE/NoPE query gains; per-head data-dependent attention-output gate. \\

\midrule
\multirow{2}{0.15\textwidth}{\raisebox{-0.35\baselineskip}{NanoChat-D12}}
& \texttt{007}
& \texttt{SSSL} to \texttt{L} attention path; masked SDPA math layers moved to Flash SDPA. \\
\cmidrule(l){2-3}
& \texttt{022/094, 031, 104, 109}
& GQA K/V projections; learnable U-net skip; Bigram Hash Embedding; MTP-2 objective. \\

\midrule
\multirow{2}{0.15\textwidth}{\raisebox{-0.35\baselineskip}{\makecell[l]{CIFAR-10\\Airbench96}}}
& \texttt{040/044/053, 059/062}
& Residual-preserved ConvGroup depth reductions; wider-shallower blocks under the accuracy gate. \\
\cmidrule(l){2-3}
& \texttt{078/081, 091/093, 090}
& Self-paced loss caching replacing the proxy model; proxy-architecture rewrites; stochastic depth on residual paths. \\

\bottomrule
\end{tabularx}
\end{table}

\begin{table}[t]
\centering
\footnotesize
\caption{Final-recipe components and trajectory evidence. Major components are code-level or protocol-level changes. Small coefficient adjustments are listed only when they are the final polish on a larger branch.}
\label{app:tab:final-recipe-components}
\setlength{\tabcolsep}{4.5pt}
\renewcommand{\arraystretch}{1.16}
\begin{tabularx}{\textwidth}{@{}
>{\raggedright\arraybackslash}p{0.15\textwidth}
>{\raggedright\arraybackslash}p{0.19\textwidth}
>{\raggedright\arraybackslash}X
>{\raggedright\arraybackslash}p{0.23\textwidth}
@{}}
\toprule
Task & Final improvement & Major developed components & Trajectory evidence \\
\midrule
Parameter Golf &
1.0810 to 1.072210, 0.81\% lower &
Score-first evaluation-time adaptation with TTT-only z-loss, loop-aware residual scaling for recurrent blocks, separate RoPE/NoPE query gains, attention-output gating, and GPTQ/Hessian calibration changes. The final optimizer edit decouples Muon warmdown cool target from warmup start. &
exp\_587 finds useful z-loss but exceeds the size cap. exp\_596 repairs the artifact and becomes legal. exp\_746 and exp\_750 refine the same head. \\
\midrule
NanoChat-D12 &
0.1618 to 0.2244, 38.7\% higher &
Attention-path rewrite from \texttt{SSSL} to \texttt{L} so all layers use Flash SDPA in the local GPU environment, expanded training-token budget under the same 90 minute cap, and a final learnable logit-bias path after \texttt{lm\_head}. &
exp\_007 identifies the backend bottleneck. exp\_020, exp\_024, and exp\_025 spend recovered wallclock on more tokens. exp\_156 adds the vocabulary-prior path. \\
\midrule
CIFAR-10 Airbench96 &
26.3560 s to 25.1464 s, 4.59\% lower &
Gate-aware speed recipe that skips most logging-only intermediate validation calls, shortens the training horizon, raises learning-rate intensity, and completes warmup earlier. Pure architecture and proxy-model speedups were tried but usually missed the 0.96 accuracy gate. &
exp\_007 and exp\_008 remove overhead and shorten the run. exp\_030 reduces evaluation overhead further. exp\_060 shows a fast near-gate miss. exp\_070 repairs the gate with 5\% warmup. \\
\bottomrule
\end{tabularx}
\end{table}

\paragraph{Failure rows as boundary evidence.}
Failed trials are not the main product, but they are part of the
feedback loop. CIFAR ineligible rows map the speed and accuracy boundary.
Parameter Golf size and evaluation failures name which proposal families
are too expensive or too large. NanoChat-D12 has few real crashes because
preflight catches common failure classes before a full run starts. A
compact memory of failure type, crash excerpt, and phase timing gives
the next agent a concrete boundary to respect rather than a vague
instruction to try something different.

\paragraph{Measurement turns proposals into evidence.}
The environment must own the metric. If the editable recipe can report
its own time, accuracy, or loss, the agent can improve the row without
improving the training run. This is why CIFAR uses shell-side timing,
NanoChat-D12 parses the training log, and Parameter Golf uses the
external evaluation path. Calibration serves the same purpose at the
starting-point level. Hardware, runtime images, offline settings, and seed
protocols change the absolute number. Running the unmodified recipe
under the same protocol makes the relative improvement meaningful.

\paragraph{Evaluator-touch audit.}
We audit the edit surface for evaluator contact. In NanoChat-D12,
visible edits touch model, training, data, optimizer, and experiment
files, with zero edits to sensitive evaluator or parser paths. In
CIFAR-10 Airbench96, all visible edits touch \texttt{airbench96.py},
with zero edits to the timing classifier or result parser. Parameter
Golf uses an older event schema, so its evaluator-touch audit relies on
archived snapshot inspection rather than edit-stream paths alone. The
audit supports measurement hardening as a design choice, not as an
ablated mechanism.

\paragraph{Concrete feedback repair cases.}
Three rows illustrate how measured feedback becomes the next edit. In
Parameter Golf, exp\_587 measured \texttt{val\_bpb}=1.072431 with
TTT-only z-loss but missed the 16 MB artifact cap by 2,056 bytes.
exp\_596 retained the same z-loss mechanism and recovered byte headroom,
turning the idea into a legal keep at 1.072251. In NanoChat-D12,
exp\_007 moved the attention path to Flash SDPA and exposed large runtime
slack. exp\_020 and exp\_024 used that slack for more training tokens,
then exp\_025 converted the same direction into the main plateau. In
CIFAR-10 Airbench96, exp\_060 was fast at 25.1650 s but missed the gate
with 0.959560 accuracy. exp\_070 kept the 42-epoch, lr=11 speed recipe
and changed warmup from 10\% to 5\%, reaching 25.1464 s with 0.960080
accuracy.

\section{Detailed final solutions and schematics}
\label{app:final-solutions}

This appendix gives a full prose description of the final recipe on each environment, lays out each developed component against the inherited starting stack, and summarizes the data flow with a schematic. The schematics were produced post hoc as explanatory figures by Claude Design and were not part of the search loop itself. Components that the closed-loop search added or rewrote are marked in teal in every schematic.

\subsection{Recipe-specific term glossary}
\label{app:recipe-glossary}

\begin{figure}[t]
\centering
\includegraphics[width=\linewidth]{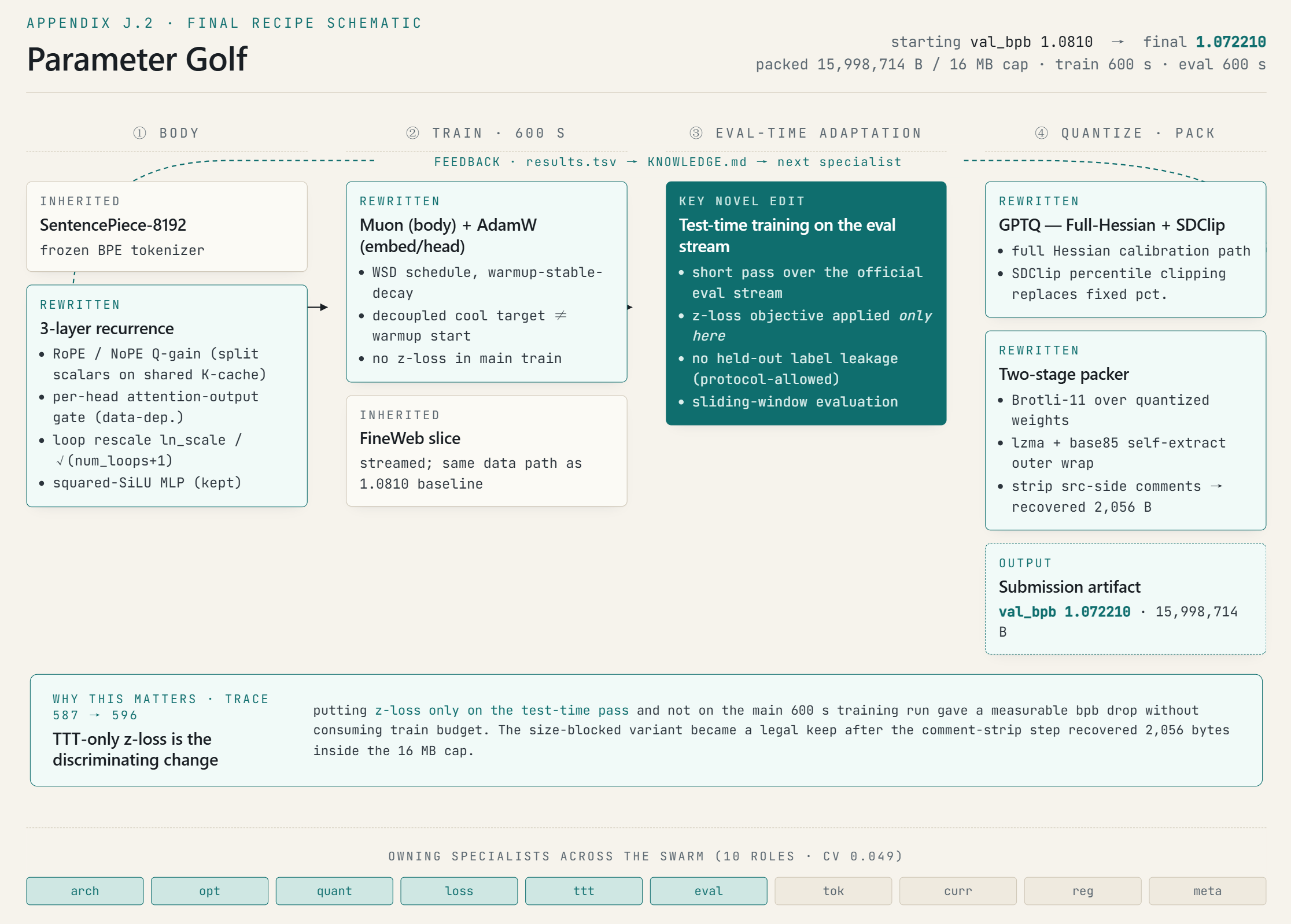}
\caption{Parameter Golf final recipe schematic. The figure summarizes inherited and rewritten components, the score-first evaluation-time adaptation path, the feedback signal that re-enters lineage, and the final artifact. \emph{Post hoc Claude Design-generated explanatory schematic; not part of the search loop.}}
\label{app:fig:pg-schematic}
\end{figure}

\paragraph{Parameter Golf.}
\emph{Evaluation-time adaptation} denotes score-first test-time updates inside the sliding-window evaluation flow: a chunk is scored under the current model before any gradient update from that chunk, and only then can that chunk train the model for later chunks. \emph{TTT-only z-loss} adds the z-loss objective only during those score-first updates, so the auxiliary objective does not consume the main train budget. \emph{Separate RoPE and NoPE query gains} apply different learnable scalar gains to rotary and non-rotary projection heads, and a \emph{per-head attention-output gate} multiplies each head's attention output by a data-dependent scalar before the residual add.

\paragraph{NanoChat-D12.}
The \emph{\texttt{SSSL} to \texttt{L} attention path} rewrites the 12 layer body from short masked-SDPA sliding-window layers mixed with longer Flash-SDPA layers into a uniform stack of long Flash-SDPA layers. \emph{Flash SDPA} is the IO-aware attention kernel used by that dispatch path. The \emph{logit-bias path} is a zero-initialized learnable vector added after \texttt{lm\_head} and trained as a vocabulary-level prior.

\paragraph{CIFAR-10 Airbench96.}
The \emph{warmup repair} reduces the warmup ratio so the schedule reaches its peak earlier, recovering the accuracy margin lost when the run is shortened.

\subsection{Parameter Golf final recipe}
\label{app:final-pg}

The starting recipe is the code path corresponding to the public 1.0810 Parameter Golf record. Its inherited stack is a SentencePiece-8192 BPE tokenizer, a transformer body with three layers carrying a long-context recurrence path stacked on parallel residual sublayers, a squared-SiLU activation in the MLP, full attention with a single rotary positional embedding for all heads, a Muon optimizer on most parameter groups with AdamW on the embeddings and language-model head, a single warmup-stable-decay schedule, GPTQ post-training quantization with a hand-tuned percentile clip, and a two-stage submission packer that first applies Brotli-11 to the quantized weights and then wraps the resulting blob plus the source code in an lzma-plus-base85 self-extracting file under the 16 MB cap. The final 1.072210 recipe keeps the high-level structure of this stack and changes a small number of components in code.

Inside attention the search adds separate query-gain scalars for the rotary and non-rotary projection heads, so that the same K-cache services both. It also adds a per-head data-dependent output gate, where the gate values are computed from the incoming residual features and used to scale each head's attention output before the residual add. Inside the recurrence path, the looped block applies a fixed rescaling of \texttt{ln\_scale\_factor} divided by the square root of \texttt{(num\_loops + 1)} to each pass, which keeps activation magnitudes stable as the loop count grows inside the same parameter budget. The MLP keeps the squared-activation pattern from the starting recipe rather than being rewritten into a gated SwiGLU. The optimizer keeps the Muon-plus-AdamW split but decouples the Muon warmdown cool target from the warmup start, so the schedule has independent control over the early and late phases. The most distinctive change happens at evaluation time. Inside the official sliding-window evaluation flow, the code scores each chunk under the current model before using that already scored chunk for a short test-time training update. The agent loop discovered that adding z-loss only to those score-first TTT updates, and not to the main training run, gave a measurable bpb drop without hitting the size cap. GPTQ is rewritten so that calibration uses a Full-Hessian path with SDClip percentile clipping rather than a fixed percentile, which changes the calibration path in a way that reduces quantization error at the same byte budget. The trace from \texttt{587} to \texttt{596} that turned the size-blocked z-loss idea into a valid keep recovered the necessary 2{,}056 bytes through source-side artifact-headroom recovery rather than by changing GPTQ calibration. Figure~\ref{app:fig:pg-schematic} traces the resulting flow.

\begin{figure}[t]
\centering
\includegraphics[width=\linewidth]{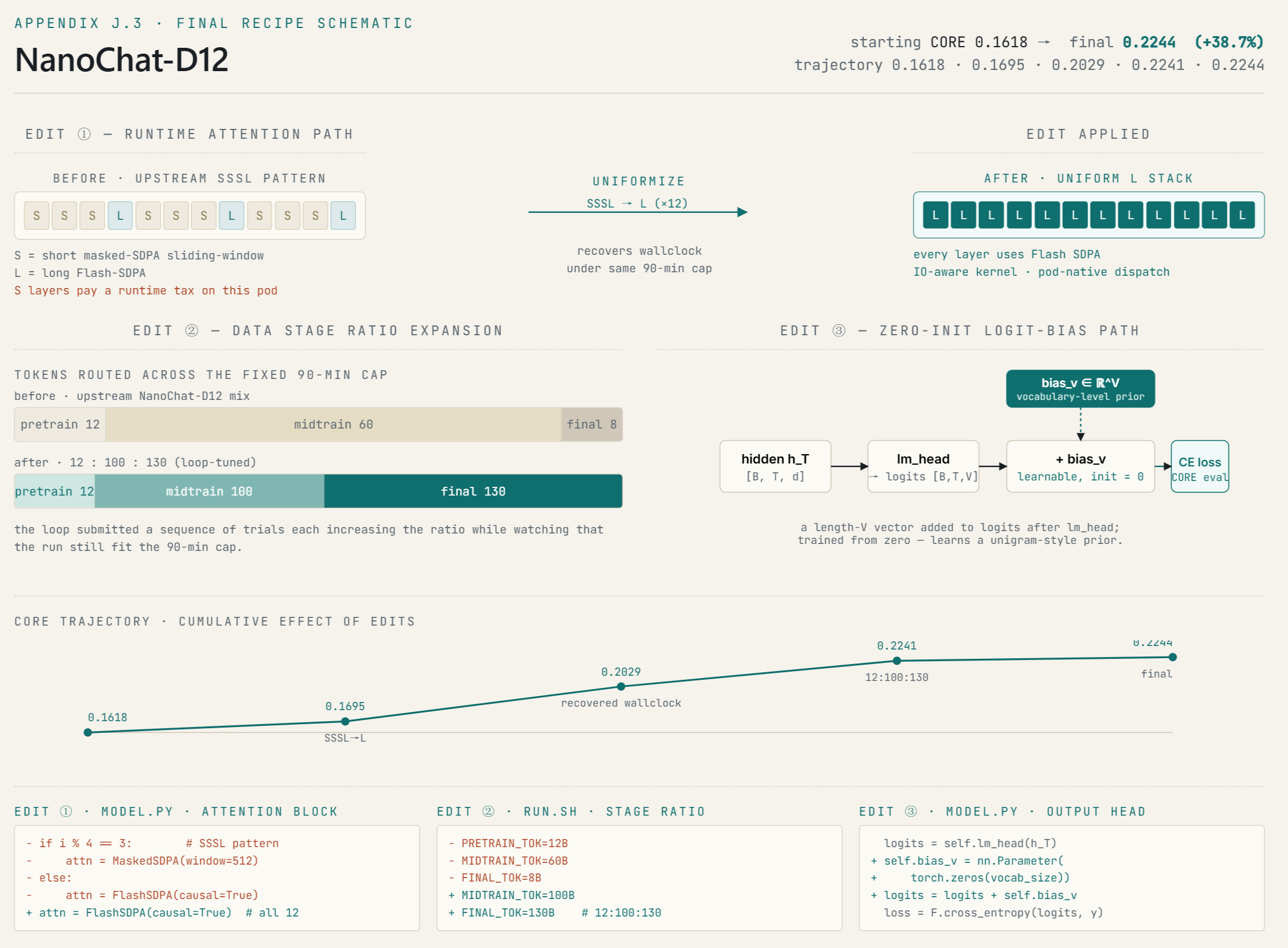}
\caption{NanoChat-D12 final recipe schematic. The figure summarizes the attention-path rewrite, data-stage ratio expansion, zero-initialized logit-bias path, and CORE trajectory. \emph{Post hoc Claude Design-generated explanatory schematic; not part of the search loop.}}
\label{app:fig:nc-schematic}
\end{figure}

\subsection{NanoChat-D12 final recipe}
\label{app:final-nc}

The starting recipe is the unmodified upstream NanoChat-D12 pretraining script at the pinned commit, calibrated to a CORE score of 0.1618 in our GPU environment. The inherited stack uses a frozen BPE tokenizer, a 12 layer transformer body that mixes short masked-SDPA sliding-window layers with longer Flash-SDPA layers in an SSSL pattern, a Muon-plus-AdamW optimizer split, a fixed 90 minute training cap, a fixed mix of pretraining-stage and midtraining-stage tokens, and a final language-model head with no learnable bias path on top of \texttt{lm\_head}. The final 0.2244 recipe keeps the optimizer and the 90 minute cap and changes three components in code.

The first change is a runtime attention-path rewrite. The 12 body layers move from the SSSL pattern to a uniform L pattern, so every layer uses Flash SDPA in the local GPU environment. This removes the runtime tax that the masked-SDPA sliding-window layers paid on this hardware and recovers a measurable amount of wallclock under the same 90 minute cap. The recovered wallclock turns into more training tokens. The recipe expands the data ratio across the pretraining, midtraining, and small final stages to roughly 12 to 100 to 130, which the loop tuned by submitting a sequence of trials that each increased the ratio while watching whether the run still fit inside the cap. The third change is a learnable logit-bias path inserted after \texttt{lm\_head} with a zero initialization, which acts as a vocabulary-level prior the model can learn during the same fixed run. CORE rises from the runtime jump to 0.1695 at the first keep, then to 0.2029 once the recovered wallclock is spent on more tokens, then to 0.2241 and 0.2244 as the data ratio and the logit-bias path are added on top. Figure~\ref{app:fig:nc-schematic} traces the resulting flow.

\begin{figure}[t]
\centering
\includegraphics[width=\linewidth]{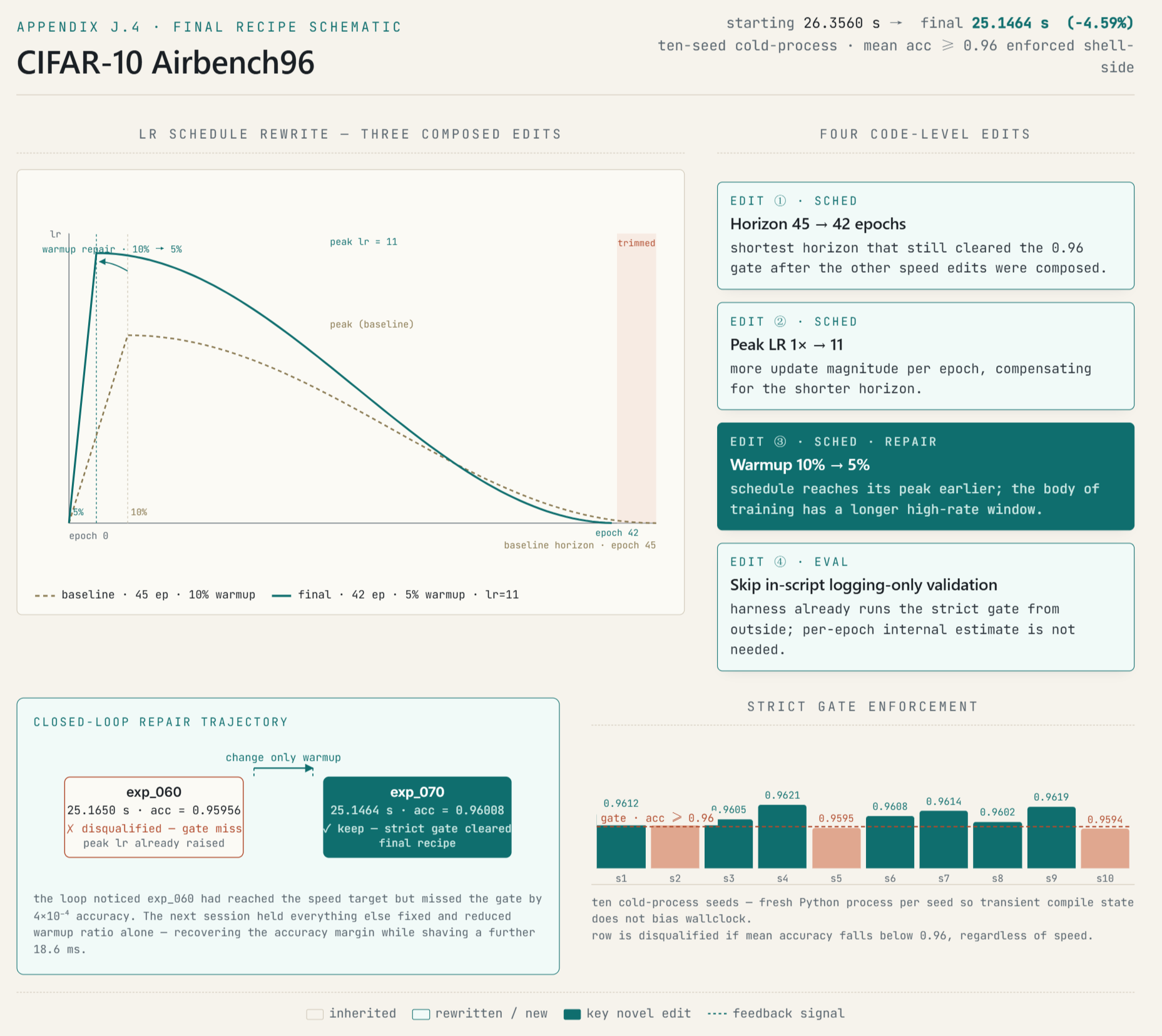}
\caption{CIFAR-10 Airbench96 final recipe schematic. The figure summarizes the schedule rewrite, four code-level edits, strict gate enforcement, and closed-loop warmup repair trajectory. \emph{Post hoc Claude Design-generated explanatory schematic; not part of the search loop.}}
\label{app:fig:cifar-schematic}
\end{figure}

\subsection{CIFAR-10 Airbench96 final recipe}
\label{app:final-cifar}

The starting recipe is the unmodified Airbench96 release, calibrated under our ten-seed cold-process protocol to a 26.3560 second mean wallclock at a strict 0.96 mean-accuracy gate. The inherited stack is a fast CIFAR-10 ConvNet, an SGD-style training loop with a linear warmup and cosine decay, a 10 percent warmup ratio, an in-script logging path that runs a small validation evaluation after every epoch, a fixed 45 epoch horizon, and a calibrated learning rate. The final 25.1464 second recipe keeps the network and the optimizer family and changes four components in code.

The first change skips most of the in-script logging-only validation calls and keeps only an end-of-training check plus an occasional intermediate one every several epochs, since the harness already runs the strict gate from outside and a per-epoch internal estimate is not needed. The second change shortens the training horizon to 42 epochs, which is the shortest horizon the loop found that still cleared the 0.96 gate after the other speed-recipe changes were composed. The third change raises the peak learning rate to 11, compensating for the shorter horizon by spending more update magnitude per epoch. The fourth change repairs the accuracy margin by reducing the warmup ratio from 10 percent to 5 percent, so the schedule reaches its peak earlier and the body of training has a longer high-rate window. The repair was triggered by exp\_060, which had already reached 25.1650 seconds with the peak learning rate raised, but missed the gate at 0.95956 mean accuracy. exp\_070 kept the same speed recipe and changed only the warmup ratio, reaching 25.1464 seconds at 0.96008 mean accuracy. Figure~\ref{app:fig:cifar-schematic} traces the resulting flow.

\section{Broader impacts and asset licenses}
\label{app:impact-licenses}

\paragraph{Broader impacts.}
The positive impact of this work is a more auditable path for empirical ML research. A closed feedback loop records hypotheses, code edits, evaluator outcomes, and failures, so follow-up work can inspect how a result was developed rather than only seeing a final recipe. It may also reduce the cost of improving small training recipes by spending bounded compute on externally verified experiments. The negative impact is that the same automation pattern could accelerate benchmark overfitting, waste compute if attached to poorly designed objectives, or optimize a harmful task more quickly when the evaluator rewards the wrong behavior. Our experiments mitigate these risks by using bounded public-style research environments, evaluator-owned scoring, no private data collection, no deployment-facing model release, and trace archives that expose both successful and failed attempts. Applying the loop to sensitive domains would require stronger human review, access control, and objective auditing than the benchmark settings used here.

\paragraph{Existing assets.}
Table~\ref{app:tab:asset-licenses} lists the external assets used by the experiments. We do not redistribute raw FineWeb/CommonCrawl text, CIFAR-10 images, upstream NanoChat evaluation datasets, Claude model weights, or third-party benchmark data. The public repository at \url{https://github.com/cxcscmu/Auto-Research-Recipes} contains the harness code, prompt templates, trace metadata, final recipes, release documentation, and pointers for users to obtain third-party assets under their own terms.

\begin{table}[ht]
\centering
\small
\caption{External assets used in the experiments and the license or terms we rely on.}
\label{app:tab:asset-licenses}
\begin{tabularx}{\linewidth}{p{0.23\linewidth}p{0.36\linewidth}X}
\toprule
Asset & Use in this paper & License or terms \\
\midrule
OpenAI Parameter Golf~\citep{openai2025parametergolf} &
Challenge harness, starting recipe, fixed task protocol, and public starting score. &
MIT License for the public repository. Challenge data are used through the task harness and are not redistributed. \\
FineWeb / CommonCrawl~\citep{penedo2024fineweb} &
Fixed language-model data source underlying the Parameter Golf task and related recipe context. &
Open Data Commons Attribution License (ODC-By) v1.0, subject to CommonCrawl Terms of Use. \\
nanochat~\citep{karpathy2025nanochat} &
NanoChat-D12 starting recipe, vendored editable Python tree, and training pipeline. &
MIT License. Upstream evaluation datasets used by nanochat retain their original terms and are not redistributed. \\
DataComp-LM / DCLM CORE~\citep{li2024datacomplm} &
CORE-style evaluation target for NanoChat-D12. &
MIT License for the framework code. Evaluation subdatasets retain their original licenses and terms. \\
cifar10-airbench~\citep{jordan2024airbench} &
CIFAR-10 Airbench96 starting recipe and speed-run structure. &
MIT License. \\
CIFAR-10~\citep{krizhevsky2009cifar} &
Image-classification data for the Airbench96 environment. &
CC BY 4.0 in UCI dataset metadata (\url{https://doi.org/10.24432/C5889J}). Raw images are not redistributed by this paper. \\
Anthropic Claude API models &
Language-agent proposal generation and post hoc Claude Design explanatory schematics. &
Anthropic Commercial API service terms (\url{https://www.anthropic.com/legal/commercial-terms}). Model weights are not accessed or redistributed. \\
\bottomrule
\end{tabularx}
\end{table}

\section{Releasable trace contents}
\label{app:traces}

A frozen run produces a \texttt{blackboard/} directory. The public repository at \url{https://github.com/cxcscmu/Auto-Research-Recipes} releases the subset needed to inspect the reported trajectories and final recipes without exposing raw runtime telemetry:

\begin{itemize}\itemsep0pt
  \item \texttt{results.tsv} --- one row per submitted trial with proposing role, hypothesis, parent \texttt{exp\_id}, status, measured score, $\Delta$ vs prior best, train / eval / total wallclock seconds, packed artifact bytes, and harness notes. Append-only.
  \item \texttt{tree.tsv} --- the same rows in preorder-sorted form with \texttt{depth} and slash-joined \texttt{path} columns so subtrees are contiguous and a single \texttt{awk} can slice an entire branch.
  \item \texttt{best.json} --- the current-best row at the moment of write. Updated atomically every time a new \texttt{keep} lands.
  \item \texttt{KNOWLEDGE.md} and \texttt{LEADERBOARD.md} --- de-identified lineage summaries and top keep rows used for audit and compact replay of the reported trajectories.
  \item \texttt{snapshots/<exp\_id>\_<role>/} --- frozen keep-time or final-recipe workdir snapshots for the reported developed recipes and controls.
  \item Prompt templates, harness code, and reproduction notes in the public repository.
\end{itemize}

The public repository intentionally omits full per-trial stdout, raw runtime event logs, full per-session rendered prompts, full submitted-trial code snapshots, scratch workdirs, and supervisor telemetry. Those files are not needed to inspect the reported score trajectory or final recipes and may contain low-level runtime accounting. The released archive is sufficient to audit the submitted trial rows, follow parent-child lineage, compare released keep/final code snapshots with the reported recipes, and reproduce the final submitted solutions after preparing the third-party task assets under their original terms.

\stopcontents[appendix]

\end{document}